\documentclass[11pt]{article}
\usepackage{cite}
\usepackage{amsmath,amsfonts,amssymb}%,calrsfs}
\usepackage[small,bf,hang]{caption}
\usepackage{slashed}

\makeatletter\let\corresponds\@undefined\makeatother
\usepackage{mathabx}

\def\hybrid{
        \topmargin -20pt
        \oddsidemargin 0pt
        \headheight 0pt \headsep 0pt
        \textwidth 6.25in % A4 paper
        \textheight 9.5in % A4 paper
        \marginparwidth .875in
        \parskip 5pt plus 1pt \jot = 1.5ex}

\hybrid

 \csname
@addtoreset\endcsname{equation}{section}

\def\moth{\mathsurround=0pt}
\newdimen\zo \zo=0pt

\def\tick{\leaders\hrule height 0.5ex depth 0pt \hskip 0.5pt}
\def\upboxfill{$\moth \setbox\zo\hbox{\tick}%
  \hskip 3pt\hbox to 0pt{$\tick$\hss}\hrulefill \hbox to 7.5pt{$\tick$\hss}$}

\def\dtick{\leaders\hrule height .34pt depth 0.5ex \hskip 0.5pt}
\def\downboxfill{$\moth \setbox\zo\hbox{\dtick}%
  \hskip 2pt\hbox to 0pt{$\dtick$\hss}\hrulefill \hbox to 2pt{$\dtick$\hss}$}

\def\bec{\begin{center}}
\def\ec{\end{center}}

\def\e{\epsilon}

\def\x{\xi}

 \def\det{{\rm det\,}}
\def\be{\begin{equation}}
\def\ee{\end{equation}}
\def\bea{\begin{eqnarray}}
\def\eea{\end{eqnarray}}
\def\ba{\begin{array}}
\def\ea{\end{array}}

\begin{document}

\begin{titlepage}
\rightline{}
\rightline{\tt MIT-CTP-4385}
\rightline{\tt  LMU-ASC~48/12}
%\rightline\today
\rightline{July 2012}
\begin{center}
\vskip 2.5cm
{\Large \bf {
Large Gauge Transformations in Double Field Theory
}}\\
\vskip 2.5cm
{\large {Olaf Hohm${}^1$ and Barton Zwiebach${}^2$}}
\vskip 1cm
{\it {${}^1$Arnold Sommerfeld Center for Theoretical Physics}}\\
{\it {Theresienstrasse 37}}\\
{\it {D-80333 Munich, Germany}}\\
olaf.hohm@physik.uni-muenchen.de
\vskip 0.7cm
{\it {${}^2$Center for Theoretical Physics}}\\
{\it {Massachusetts Institute of Technology}}\\
{\it {Cambridge, MA 02139, USA}}\\
zwiebach@mit.edu

\vskip 1.5cm
{\bf Abstract}
\end{center}

\vskip 0.4cm

\noindent
\begin{narrower}

Finite gauge transformations in double field theory can be 
defined by  the exponential of generalized
Lie derivatives.  
We  interpret  these transformations as  
 `generalized coordinate transformations'
in the doubled space by proposing and testing 
a formula that writes large transformations in terms 
of derivatives of the coordinate maps.
Successive generalized coordinate transformations give
a generalized coordinate transformation that differs from
the direct  composition of the original two.
Instead, it is constructed using the Courant bracket.  
These transformations form a group when acting on fields 
but, intriguingly, do not associate when acting on coordinates.

\end{narrower}

\end{titlepage}

\newpage

\tableofcontents

\section{Introduction}
Double field theory is a spacetime description of the massless sector of closed string theory 
that makes T-duality 
manifest 
by doubling the coordinates. In addition to the usual spacetime coordinates
$x^{i}$, $i=0,\ldots, D-1$,
there are 
dual `winding' coordinates $\tilde{x}_{i}$, which together with the $x^i$   
combine into 
coordinates $X^{M}=(\tilde{x}_{i},x^{i})$ transforming in the fundamental representation 
of the T-duality group $O(D,D)$. This theory has been formulated in \cite{Hull:2009mi,Hull:2009zb,Hohm:2010jy,
Hohm:2010pp}. Earlier important work can be found in \cite{Siegel:1993th,Tseytlin:1990nb,Duff:1989tf}
and further developments have been discussed in \cite{Hohm:2010xe,Kwak:2010ew,Hohm:2011gs,Hohm:2011dz,Hohm:2011ex,Hohm:2011zr,arXiv1108.4937,arXiv1111.7293,Hohm:2011si,Hillmann:2009ci,Berman:2010is,
West:2010ev,Jeon:2010rw,Jeon:2011cn,Jeon:2011vx,Schulz:2011ye,Copland:2011yh,
Thompson:2011uw,Albertsson:2011ux,Andriot:2011uh,aldazabal-geiss,grana-marques,Coimbra:2011nw,Vaisman:2012ke,
Andriot:2012wx,Dibitetto:2012rk,Kikuchi:2012za,Malek:2012pw}. 

There are various formulations of double field theory. This paper
uses the generalized 
metric formulation~\cite{Hohm:2010pp},  
in which the fundamental dynamical field is the $O(D,D)$ matrix 
 \be\label{IntroH}
  {\cal H}_{MN} \ = \  \begin{pmatrix}    g^{ij} & -g^{ik}b_{kj}\\[0.5ex]
  b_{ik}g^{kj} & g_{ij}-b_{ik}g^{kl}b_{lj}\end{pmatrix}\;,
 \ee
that unifies the spacetime metric $g_{ij}$ and the Kalb-Ramond 
two-form  
 $b_{ij}$ and that 
transforms covariantly under $O(D,D)$. In addition, the theory features the dilaton $d$, 
which is a scalar under $O(D,D)$. This dilaton field  is a spacetime density and is related to the scalar dilaton 
$\phi$ through the field redefinition 
$e^{-2d}=\sqrt{-g}e^{-2\phi}$. 
The double field theory action can be written in terms of a generalized curvature scalar 
${\cal R}$ that is a function of ${\cal H}$ and $d$ \cite{Hohm:2010pp}, 
 \be\label{DFTaction}
  S_{\rm DFT} \ = \  \int dxd\tilde{x}\,e^{-2d}\,{\cal R}({\cal H},d)\;.
 \ee
This curvature scalar is a manifestly $O(D,D)$ invariant expression in terms of 
${\cal H}$, $d$ and `doubled' derivatives $\partial_{M}=(\tilde{\partial}^{i},\partial_i)$, and 
so the $O(D,D)$ invariance of (\ref{DFTaction}) is manifest. 
This theory also features a gauge invariance whose infinitesimal transformations are 
parametrized by an $O(D,D)$ vector parameter $\zeta^{M}=(\tilde{\zeta}_{i},\zeta^{i})$ that 
combines the diffeomorphism parameter $\zeta^{i}$ and the $b$ field gauge parameter 
$\tilde{\zeta}_i$. It acts on the physical fields as 
 \be\label{gaugevar}
 \begin{split}
  \delta_{\zeta}{\cal H}_{MN} \ &= \ 
  \zeta^{P}\partial_{P}{\cal H}_{MN}+\big(\partial_{M}\zeta^{P}-\partial^{P}\zeta_{M}\big)
  {\cal H}_{PN}+\big(\partial_{N}\zeta^{P}-\partial^{P}\zeta_{N}\big)
  {\cal H}_{MP}\;, \\
  \delta_{\zeta}d \ &= \ \zeta^{M}\partial_M d-\frac{1}{2}\partial_{M}\zeta^{M}\;. 
 \end{split}
 \ee  
 We may define a generalized Lie derivative 
$\widehat{\cal L}_\zeta$ acting on 
$O(D,D)$ tensors with 
arbitrary index structure.  For the generalized metric the above gauge transformation is in fact the generalized
Lie derivative:   
$\delta_{\zeta}{\cal H}_{MN} =  \widehat{\cal L}_{\zeta}{\cal H}_{MN}$.
Under these variations ${\cal R}$ 
transforms as a generalized scalar, 
$ \delta_{\zeta}{\cal R}  =  \zeta^{M}\partial_{M}{\cal R}$,
from which the gauge invariance of (\ref{DFTaction}) immediately follows. More precisely, in order to 
verify this invariance
the following `strong constraint' is required:  
  \be\label{STRONG}  
  \partial^{M}\partial_{M} \ \equiv \  
  \eta^{MN}\partial_{M}\partial_{N}  \ = \ 0
   \;, \quad \hbox{with} \qquad 
  \eta^{MN} \ = \ \begin{pmatrix}    0 & {\bf 1} \\[0.5ex]
  {\bf 1} & 0 \end{pmatrix}\;.
   \ee
The above constraint 
must hold when acting on arbitrary fields and parameters and all their products (so that $\partial^M\partial_M A = 0$  
 and $\partial^M A\, \partial_M B=0$
for any fields or parameters $A$ and $B$).  Here $\eta_{MN}$ denotes the $O(D,D)$ invariant metric. 
This constraint actually implies  that one can always find an $O(D,D)$ rotation into a T-duality frame in which the coordinates 
depend only, say, on the $x^i$. 
 
Satisfying this constraint by 
setting $\tilde{\partial}^i=0$, the action (\ref{DFTaction}) reduces to 
the standard low-energy effective action for the NS-NS sector of closed string theory.  
Moreover, the gauge variations (\ref{gaugevar}) reduce for the components in (\ref{IntroH})
precisely to the standard (infinitesimal) general coordinate transformations and $b$ field gauge transformations. 
We stress that the gauge transformations (\ref{gaugevar})
are not infinitesimal diffeomorphisms on the doubled space,   
because 
they do not close according to the Lie bracket 
but rather according to  
the `C-bracket'~\cite{Hull:2009zb,Hohm:2010pp,Siegel:1993th}, 
 \be\label{Cbracket}
  \big[\delta_{\zeta_1},\delta_{\zeta_2}\big] \ =
   \ -\delta_{\,[\zeta_1,\zeta_2]_c} 
  \;, \qquad
   \bigl[ \zeta_1,\zeta_2\bigr]_c^M  
   \ \equiv
   \ \zeta_{1}^{N}\partial_{N}\zeta_{2}^M -\frac{1}{2}\, 
    \zeta_{1N}\partial^{M}\zeta_{2}^N
   -(1\leftrightarrow 2)\;, 
 \ee
which is the $O(D,D)$ covariant extension of the Courant bracket of generalized geometry. 

In this paper we will investigate the finite or large gauge transformations 
corresponding to the infinitesimal variations (\ref{gaugevar}). Since 
these gauge variations do not represent infinitesimal diffeomorphisms 
of the doubled 
space we cannot  
resort  to Gauss and Riemann and postulate 
the usual coordinate transformation rules of vectors and one-forms. 
In fact, inspection of  (\ref{gaugevar}) shows that each index appears to 
be some `hybrid' between covariant and contravariant indices.
It is thus not clear how finite transformations can be consistently defined.

We find, however, that it is possible to view finite gauge
transformations as arising from some suitably defined
`generalized coordinate transformations'.  
We introduce such
coordinate transformations   
with the features that are expected from the infinitesimal 
gauge transformations. This implies 
that they do not satisfy all the properties of diffeomorphisms.  
For instance,  two successive diffeomorphisms give a third diffeomorphism that is simply defined by 
direct 
 composition of the first two. 
Two successive  `generalized coordinate transformations' 
also result in a generalized coordinate transformation, but the
resulting transformation is not obtained by the direct composition of the two maps.   
This is the group 
manifestation of the fact that the 
gauge algebra is governed by the Courant bracket (\ref{Cbracket}) rather than the Lie bracket.

Given a generalized coordinate transformation 
$X\rightarrow X^{\prime} = f(X)$, we propose the following
 associated
transformation for an 
$O(D,D)$ vector $A_{M}$: 
 \be\label{finiteFIntro}
   A_{M}^{\prime}(X^{\prime}) \ = \ {\cal F}_M{}^N   A_N(X)\,,
   \ee
where the matrix ${\cal F}$ is defined by
   \be\label{calFdefIntro} \phantom{\Biggl(}
 {\cal F}_M{}^N\ \equiv \   {1\over 2} 
  \Bigl(   \frac{\partial X^{P}}{\partial X^{\prime M}}\,
  \frac{\partial X^{\prime}_P}{\partial X_{N}}
  + \frac{\partial X'_{ M}}{\partial X_P}\,
  \frac{\partial X^N}{\partial X^{\prime P}}\Bigr)
  \,. 
 \ee
Here 
the indices on coordinates are raised and lowered with the $O(D,D)$ invariant metric, 
$X_{M}=\eta_{MN}X^{N}=(x^{i},\tilde{x}_i)$, etc. 
More generally, a tensor with an arbitrary number of $O(D,D)$ indices transforms `tensorially', 
with each index rotated by the matrix ${\cal F}$.   We show that
${\cal F}$ is in fact an $O(D,D)$ matrix.  
In ordinary geometry we would simply have ${\cal F}_M{}^N={\partial X^N\over \partial X'^M}$. 
In double field theory 
the dilaton $d$ provides the scalar density $\exp (-2d)$. We give the transformation
law for this density under large
coordinate transformations in (\ref{densitytrans}).

We will show that
the transformation rule in (\ref{finiteFIntro}) and (\ref{calFdefIntro}) implies the infinitesimal transformations~(\ref{gaugevar}) 
when we set $X' = X - \zeta (X)$. 
We have also verified that this transformation
satisfies the following consistency requirements:
It implies the usual formulae  
for 
coordinate transformations that 
transform only the $x^{i}$ or only
the $\tilde{x}_i$.  It leaves the $O(D,D)$ invariant metric in  (\ref{STRONG}) invariant, i.e., 
this metric takes the same \textit{constant} form in all coordinate systems, something required in double field theory but
inconsistent in conventional differential geometry.  
Moreover, the strong constraint (\ref{STRONG}) in one coordinate system implies the strong constraint 
in all other coordinate systems.

As mentioned above, the generalized coordinate transformations do not compose like ordinary diffeomorphisms. 
In order to elucidate this point, it is useful to introduce an alternative form of the finite gauge transformations.  
The rule (\ref{finiteFIntro}) defines the transformed tensor by giving its transformed components at the 
\textit{transformed} point $X^{\prime}$. 
As in general relativity, this can be seen as a \textit{passive} transformation, but 
it is useful to also have an \textit{active} form of the gauge transformations which transforms 
the field components, but not the coordinates. For general relativity this problem has been discussed in the 
literature, see, e.g., \cite{Bruni:1996im,Abramo:1997hu}, where it is found that gauge transformations connected to the 
identity can be realized as an exponential of the Lie derivative. 
Thus, given an ordinary vector field $A_m(x)$ we have the 
transformed field $A'_m(x)$ given by
\be
A'_m (x) \ = \ e^{{\cal L}_\xi}   \, A_m(x) \,, 
\ee
where ${\cal L}_\xi$ is the Lie derivative in the representation
appropriate for a vector, and all fields and parameters depend on 
$x$.  It can be shown that this transformation is induced by the following diffeomorphism
\be
x'^m  \ = \  e^{-\xi^k \partial_k} \, x^m \,.
\ee

In double field theory we can follow the above strategy. 
Even though the Courant bracket does not define a Lie algebra, generalized
Lie derivatives define 
a Lie algebra
under commutators.  
We can therefore 
realize a finite gauge transformation 
by exponentiating the generalized Lie derivative:
 \be\label{EXpLie}
  A_{M}^{\prime}(X) \ = \ e^{\widehat{\cal L}_{\xi}} \,A_{M}(X)\;, 
 \ee  
where 
all fields and parameters 
depend on $X$.  
If finite gauge transformations are defined this way it 
is simple to use the Baker-Campbell-Hausdorff  
formula to show that the field transformations form a group and 
compose according to the Courant bracket.  
Our key technical result is the determination of the
generalized coordinate transformation
\be
X'^M \ = \  e^{-\Theta^K(\xi)\partial_K}  X^M \,, ~~~~
\Theta^K(\xi) \, \equiv  \ \xi^K   + {\cal O} (\xi^3)\,,
\ee
so that (\ref{finiteFIntro}) and (\ref{calFdefIntro})  
lead   
to the transformation (\ref{EXpLie}), 
at least to ${\cal O}(\xi^4)$.  
The composition rule for 
generalized coordinate transformations, 
calculable from 
the definition
(\ref{EXpLie}),  will be verified
explicitly with 
${\cal F}$ expanded  
to quadratic order in~$\xi$. 
We note in passing that while the exponential (\ref{EXpLie}) only makes sense  
for gauge transformations connected to the identity, the 
generalized coordinate transformations may be applicable more 
generally.

Even though the composition rule 
is non-standard, we are intrigued that 
the simple 
generalization of conventional tensor 
transformations 
given by (\ref{finiteFIntro}) and
(\ref{calFdefIntro}) exists, 
seems to pass all consistency checks, and 
is, plausibly, 
the unique form compatible with (\ref{EXpLie}). 
Surprisingly, while generalized 
coordinate transformations form a group when acting on fields, 
they do not satisfy associativity at the 
level of coordinate maps. Further discussion of this result and 
other open questions can
be found in the concluding section.

\section{Finite gauge transformations} \label{Fingautra}

In this section we 
propose finite gauge transformations
for double field theory.  These transformations are induced by
(and written in terms of) generalized coordinate transformations. 
We begin by discussing these coordinate transformations
and compute the derivatives of the maps using a  
simple parameterization. 
We show that the strong
constraint is preserved by these coordinate transformations and  
that applying the transformation rule to $\partial_M$ is consistent with the chain rule. 
Finally,  $\eta_{MN}$ is an invariant tensor so 
that ${\cal F}$ is actually an $O(D,D)$ matrix.

\subsection{Coordinate transformations  
and strong constraint}\setcounter{equation}{0}

In this subsection we describe some 
generalized coordinate transformations of the doubled coordinates.
We will use throughout 
section~\ref{Fingautra} and \ref{spegautraandoddtra}  
 -- but not in the rest of the paper -- a parameterization with a parameter $\zeta^{M}(X)$ and new coordinates $X'$  given by
the exact relation 
 \be\label{xi}
  X^{\prime M} \ = \ X^{M}-\zeta^{M}(X)\; .
 \ee 
It follows by differentiation that 
\be\label{coordDiff}
{\partial X'^Q \over \partial X^P ~} \ = \  \delta_P{}^Q \ - \  \partial_P \zeta^Q \,,  
\ee
and in matrix notation we write this as  
\be
\label{px'px}
\Bigl({\partial X' \over \partial X}\Bigr)_P^{~~Q}\  \equiv \
{\partial X'^Q \over \partial X^P ~} \ = \  \bigl( {\bf 1} -  a\bigr)_P{}^Q  \,,
~~~\hbox{with}  \qquad  a_P{}^{Q} \equiv  \ \partial_P \zeta^Q\,.
\ee
Note that when  
 representing coordinate derivatives as matrices we will always associate the first index
(row index) with the coordinate in the denominator 
and the second index (column index) with the coordinate in the
numerator. The matrix inverse provides us with the other derivatives
 \be
 \label{pxpx'}
  \Bigl( \frac{\partial X}{\partial X'}\Bigr)_{M}^{~~P} \ \equiv \ \frac{\partial X^{P}}{\partial X^{\prime M}} \ =  \ \Bigl({1\over 1-a} 
  \Bigr)_{M}{}^P \ =  \ \Bigl(
 1+ a + a^2 + a^3 + \ldots \Bigr)_{M}{}^P    \;,
  \ee
  or, more explicitly, 
   \be
   \label{pxpxprimeexp}
  \frac{\partial X^{P}}{\partial X^{\prime M}} \ =  \ \delta_{M}{}^{P}+\partial_{M}\zeta^{P} +\, \partial_{M}\zeta^{L}\partial_{L}\zeta^{P} \,
  +\, \partial_{M}\zeta^{L}\partial_{L}\zeta^{R} \partial_{R}\zeta^{P}  + {\cal O}(\zeta ^4)  \;. 
   \ee

Let us now consider the strong constraint (\ref{STRONG}).  In this setup with
large coordinate transformations we assume that $\zeta^M$ as
well as all $X$-dependent fields satisfy the strong constraint:
\be
\label{scdef}
\eta^{MN} \partial_N A\, \partial_M B \ \equiv  \ \partial^M A\,  \partial_M B \ = \ 0 \,,     ~~~~\partial^P\partial_P A \ = \ 0 \,.
\ee
In the above, $A(X)$ and $B(X)$ can be $\zeta^M$ or
any field of the theory, like the dilaton or the generalized metric.
The strong constraint implies that the product $a^t a $ vanishes:
 \be
  a^t a \ = \ 0 \,. 
 \ee
 Indeed,
 \be
 0 \ = \ \partial_P 
 \zeta_M\,  \partial^P \zeta_N  \ = \ a_{PM}\, a^P{}_N \ =  \  
 (a^t)_{MP}\,  a^P{}_N \ = \  (a^t a)_{MN} \,.
 \ee
 
We claim that if the strong constraint holds for all fields and parameters in coordinate 
system $X$ it will then hold for coordinate system $X'$. 
We begin by proving the following lemma in which two
functions $A$ and $B$ of $X$ are differentiated with 
mixed-type  
derivatives:
 \be\label{stronglemma}
  \partial^{M\prime}A\, \partial_{M}B \ = \ 0\;, \qquad 
  \partial^{M\prime}\,\partial_{M}\,A \ = \ 0\;.
 \ee
To see this we note that a primed derivative,  with 
the help of (\ref{pxpxprimeexp}), can be written as 
\be
 \begin{split}
\partial'^M \ \equiv \      
\eta^{MN}\partial_{N}^{\prime} \ &= \ \eta^{MN}\frac{\partial X^{P}}{\partial X^{\prime N}}
  \partial_{P} \\
  \ &= \ \eta^{MN}\big(\delta_{N}{}^{P}+\partial_{N}\zeta^{P}
  +\partial_{N}\zeta^{K}\partial_{K}\zeta^{P}
  +  \partial_{N}\zeta^{K}\partial_{K}\zeta^{Q}\partial_Q \zeta^P +\cdots  \big)
  \partial_{P} \\[0.3ex]
  \ &= \ \big(\eta^{MP}+\partial^{M}\zeta^{P}
  +\partial^{M} \zeta^{K}\partial_{K}\zeta^{P}+
  \partial^{M}\zeta^{K}\partial_{K}\zeta^{Q}\partial_Q \zeta^P
  + \cdots  \big)
  \partial_{P}   \\[0.3ex]
  \ &= \ \partial^{M}+  \partial^{M}\zeta^{K}  ( \delta_K{}^{P} 
  +  \partial_{K}\zeta^{P}
  +    \partial_{K}\zeta^{Q}\partial_Q \zeta^P +\cdots  \big)
  \partial_{P}   \;. 
 \end{split} 
 \ee
We see that structurally this takes the form
\be\label{primetrans}
\partial'^M  \ = \  \partial^{M}+  (\partial^{M}\zeta^{K}) \, {\cal U}_K{}^{P} 
\,  \partial_{P}  \,, 
\ee
where ${\cal U}$ is a matrix function of $X$ whose expression
is in fact not important. 
The lemmas now follow easily:
\be
 \partial'^{M}A\, \partial_{M}B \ = \ 
 \Bigl( \partial^{M} A +  (\partial^{M}\zeta^{K}) \, {\cal U}_K{}^{P} 
\,  \partial_{P} A \Bigr) \partial_{M}B  \ = \ 0 \,,
\ee
by use of the strong constraint as in (\ref{scdef}).
Similarly,  
\be
 \partial'^{M}\, \partial_{M}A \ = \ 
 \Bigl( \partial^{M}  +  (\partial^{M}\zeta^{K}) \, {\cal U}_K{}^{P} 
\,  \partial_{P}  \Bigr) \partial_{M}A  \ = \ 0 \,.
\ee
Using the lemma and 
(\ref{primetrans}) it now follows that 
\be
\label{scxprime}
\partial'^M A \,  \partial'_M B  \ = \ 0 \,, ~~~~  
\partial'^M  \partial'_M A  \ = \ 0 \,.
\ee
These 
can be viewed as the statement that the strong constraint
holds in the primed coordinates.

\subsection{Large gauge transformations}

For a scalar $S(X)$ the  coordinate transformation
will be taken to be the usual one,   
 \be\label{scalar}
  S^{\prime}(X^{\prime}) \ = \ S(X)\;.
 \ee
It then follows to first order in $\zeta$ that 
 \be
  S^{\prime}(X)-\zeta^{M}\partial_{M}S \ = \ S(X) 
  \quad \Rightarrow \quad
  \delta_{\zeta}S 
  \ \equiv   
   \ S^{\prime}(X)-S(X) \ = \ \zeta^{M}\partial_{M}S\;.
 \ee
 
For a generalized
vector $A_{M}$ we need a transformation rule that acts on it 
like for 
a one-form and a vector simultaneously. 
Indeed, here our main clue is the infinitesimal transformation
 \be
 \label{transgenvecfirstorder}
  \delta A_{M} \ \equiv  \ A_{M}^{\prime}(X)-A_{M}(X) \ =
  \ \widehat {\cal L}_\zeta  A_M \ = \  \ \zeta^{P}\partial_{P}A_{M}
  +\big(\partial_{M}\zeta^{N}-\partial^{N}\zeta_{M}\big)A_{N}\;.   
 \ee     
This must be reproduced by the formula we propose once the
parameter $\zeta$ is taken to be small. 
We propose the transformation 
 \be\label{delAM}
  A_{M}^{\prime}(X^{\prime}) \ = \ {1\over 2} 
  \Bigl(   \frac{\partial X^{P}}{\partial X^{\prime M}}\,
  \frac{\partial X^{\prime}_P}{\partial X_{N}}
  +  \frac{\partial X'_{ M}}{\partial X_P}\,
  \frac{\partial X^N}{\partial X^{\prime P}}\Bigr) \, A_{N}(X)\;. 
 \ee
In here we have defined $X_N \equiv  \eta_{NM} X^M$ and
$X'_N \equiv  \eta_{NM} X'^M$. 
Expanding to first order in $\zeta$ we find 
with (\ref{coordDiff}) and (\ref{pxpxprimeexp})
 \be
 \begin{split}
  A_{M}^{\prime}(X)-\zeta^{P}\partial_{P}A_{M}(X) \ &= \ 
 {1\over 2} \Bigl(     
 \big(\delta_{M}{}^{P}+\partial_{M}\zeta^{P}\big)
 \big(\delta^{N}{}_{P}-\partial^{N}\zeta_{P}\big) \\
 &  \quad +   \big(\delta^P{}_{M}-\partial^{P}\zeta_{M}\big)
 \big(\delta_{P}{}^{N}+\partial_{P}\zeta^{N}\big) \Bigr) A_{N}(X) \\
 \ &= \ 
 {1\over 2} \,\Bigl( 2 \delta_M{}^N  +   2 \partial_M \zeta^N 
 -    2 \partial^N \zeta_M 
  \Bigr) A_{N}(X) + {\cal O}(\zeta^2) \\
 \ &= \   A_M(X) + \big(  \partial_M \zeta^N 
 -   \partial^N \zeta_M \big) A_{N}(X) + {\cal O}(\zeta^2)\,,
\end{split}
 \ee
which indeed reproduces (\ref{transgenvecfirstorder}).   The transformation (\ref{delAM}) is not fully determined by the constraint
that the
infinitesimal transformations arise correctly.  A number of options allow for this result.  Other consistency checks 
appear to   
select (\ref{delAM})
as the only possible choice, as we will discuss in this and the following section.

Before we proceed with the analysis of the transformation (\ref{delAM})
we introduce some notation. 
We write
 \be\label{finiteF}
   A_{M}^{\prime}(X^{\prime}) \ = \ {\cal F}_M{}^N   A_N(X)\,,
   \ee
   where the matrix ${\cal F}$ is defined by
   \be
   \label{calFdef} \phantom{\Biggl(}
 {\cal F}_M{}^N\ \equiv \   {1\over 2} 
  \Bigl(   \frac{\partial X^{P}}{\partial X^{\prime M}}\,
  \frac{\partial X^{\prime}_P}{\partial X_{N}}
  + \frac{\partial X'_{ M}}{\partial X_P}\,
  \frac{\partial X^N}{\partial X^{\prime P}}\Bigr)
  \,. 
 \ee
More generally, any $O(D,D)$ tensor we require to transform under generalized coordinate transformations 
such that each index is rotated by the matrix ${\cal F}_M{}^{N}$. 

Double field theory also requires the definition of a scalar density.
The transformation (\ref{gaugevar}) of the dilation $d$ implies that 
\be
\delta_\zeta  e^{-2d} \ = \  \partial_M \bigl( \zeta^M  e^{-2d} \bigr) \,. 
\ee
This is the infinitesimal transformation of a scalar density, and it is the same transformation
that we have in ordinary differential geometry.  Thus, the 
finite gauge transformation of this density must be given by 
 \be\label{densitytrans}
  e^{-2\,d'(X')} \ = \ \Big| \det\frac{\partial X}{\partial X'} \Big| \, e^{-2\,d(X)}\;.
 \ee
Of course, using (\ref{pxpxprimeexp}) and expanding this to first order in $\zeta$ it is easily 
seen that the variation $\delta_{\zeta} d=d'(X)-d(X)$ coincides with that given in (\ref{gaugevar}). 
Further exploration of the consistency of (\ref{densitytrans}) will be discussed in 
sections~\ref{ordscaandvec},  \ref{genargcomp}, and \ref{testingcomposition}.

The transformation (\ref{delAM}) can be expanded to all orders
in $\zeta$.  
 In the matrix notation we have used for coordinate derivatives we have
 \be  
 \label{inmatrixF}  \phantom{\Biggl(} 
 {\cal F}_M{}^N (X' , X) \  = \   {1\over 2} \,
  \Bigl( \,  \frac{\partial X}{\partial X'}\,
  \Bigl(\frac{\partial X^{\prime}}{\partial X}\Bigr)^t
  +    \Bigl(\frac{\partial X'}{\partial X}\Bigr)^t\,
  \frac{\partial X}{\partial X'}\, \Bigr)_M^{~~N}  \,. 
 \ee
We have added the coordinate arguments in a specific order:
the first input is the new coordinate and the second input is the old
coordinate. We will only use those arguments when needed explicitly.
In index-free notation we write 
\be  
 \label{inmatrixFvm}  \phantom{\Biggl(} 
 {\cal F} (X' , X) \  = \   {1\over 2} \,
  \Bigl( \,  \frac{\partial X}{\partial X'}\,
  \Bigl(\frac{\partial X^{\prime}}{\partial X}\Bigr)^t
  +    \Bigl(\frac{\partial X'}{\partial X}\Bigr)^t\,
  \frac{\partial X}{\partial X'}\, \Bigr)  \,. 
 \ee
Note that ${\cal F}$ is in fact an {\em anticommutator} of partial derivatives:
\be  
 \label{inmatrixFin}
 {\cal F}(X', X) \ =  \   {1\over 2} \,\Bigl\{ \, 
   \frac{\partial X}{\partial X'}\, , \ 
  \Bigl(\frac{\partial X^{\prime}}{\partial X}\Bigr)^t\, \Bigr\} \,. 
 \ee
 
 Using our expansions (\ref{px'px}) and (\ref{pxpx'}) we immediately
 write
  \be
   \label{calFdefcalc}
 \begin{split}
 {\cal F}_M{}^N\ = 
 \ & \, {1\over 2} 
  \Bigl(  \Bigl({1\over 1-a} \Bigr) \, (1-a^t)
  + (1-a^t)   \Bigl({1\over 1-a} \Bigr) \Bigr)_M{}^{N} \\
= \ & \, {1\over 2} \,
  \Bigl( \,~1+(a-a^t)+(a^2-aa^t)+(a^3-a^2a^t)+\cdots   \\
   &\quad+ 1-a^t+a+a^2+a^3+ \cdots  \Bigr)_M{}^{N}\,,
\end{split}
 \ee
 where we used the strong constraint $a^t a =0$ and expanded in the last equation.  
  We also note that by the strong constraint between $\zeta$ and any field it follows that 
 if $A$ satisfies the strong constraint, so does $A'$ defined by (\ref{delAM}). 
 Combining these terms gives us the result
\be\label{Fpert}  
{\cal F} \ = \    1 + a - a^t + \sum_{n=2}^\infty  \bigl( a^n  - 
{1\over 2} a^{n-1} a^t \bigr)   \,. 
\ee 
Let us finally note that for $aa^t=0$ the two lines in the second equation 
of (\ref{calFdefcalc}) are equal, which in turn means that the two terms 
in the definition of ${\cal F}$ coincide, and so (\ref{inmatrixFvm}) 
reduces to one term, 
 \be\label{aatzero}
  aa^t \ = \ 0 \quad\Rightarrow \quad  
   {\cal F} (X' , X) \  = \   \frac{\partial X}{\partial X'}\,
  \Bigl(\frac{\partial X^{\prime}}{\partial X}\Bigr)^t \ = \ 
     \Bigl(\frac{\partial X'}{\partial X}\Bigr)^t\,
  \frac{\partial X}{\partial X'}\;.
 \ee
Although this does not hold in general, it does hold for a few special cases that 
we inspect in section~\ref{spegautraandoddtra}.  
 
 We now perform a basic consistency check.  We should be able
 to use the transformation (\ref{delAM}) for partial derivatives, which
 also have an index down.  Therefore, we must have
  \be\label{delAMpd}
  \partial'_{M} \ = \ {1\over 2} 
  \Bigl(   \frac{\partial X^{P}}{\partial X^{\prime M}}\,
  \frac{\partial X^{\prime}_P}{\partial X_{N}}
  +  \frac{\partial X'_{ M}}{\partial X_P}\,
  \frac{\partial X^N}{\partial X^{\prime P}}\Bigr) \, \partial_{N}\;. 
 \ee
 On the other hand, partial derivatives must also transform 
 with the chain rule  
 \be\label{Sstep1}
  \partial_{M}^{\prime}\ = \ \frac{\partial X^{N}}{\partial X^{\prime M}}\frac{\partial}{\partial X^{N}}
  \ = \ \frac{\partial X^{N}}{\partial X^{\prime M}} \partial_{N}\;.
 \ee
The two expressions are consistent thanks to the strong constraint.
For this note that the first expression
can be written as
 \be\label{delAMpdII}
  \partial'_{M} \ = \ {1\over 2} 
    \frac{\partial X^{P}}{\partial X^{\prime M}}\,
 (\delta^N{}_P - \partial^N \zeta_P) \partial_N 
  +  {1\over 2} (\delta^P{}_M  - \partial^P \zeta_M) \, \partial'_P \;. 
 \ee
By the lemmas (\ref{stronglemma}) the term $(\partial^P \zeta_M)\partial'_P$ vanishes acting on any function.  Moreover,
the term $(\partial^N\zeta_P) \partial_N$ 
also vanishes. 
Bringing the right-most non-vanishing term to the left-hand side, we
have
 \be\label{delAMpdIII}
 {1\over 2} \,  \partial'_{M} \ = \ {1\over 2} 
    \frac{\partial X^{P}}{\partial X^{\prime M}}\, \partial_P \,,
  \ee
showing that the usual transformation of derivatives is consistent
with (\ref{delAMpd}).

Our final check here is that the metric $\eta_{MN}$ 
is an invariant tensor.  For this we must have
\be
\label{invtensoreta}
\eta_{MN}  \ = \  {\cal F}_M{}^R \,{\cal F}_N{}^S \, \eta_{RS} \,. 
\ee
This equation states that ${\cal F}_M{}^N$ {\em is in fact} an
$O(D,D)$ matrix.\footnote{It should be noted, however, 
that we cannot think of the generalized coordinate transformations as local 
$O(D,D)$ transformations with an $X$-dependent $O(D,D)$ matrix $h={\cal F}(X)$. The 
reason is that in the transformation of the argument we would need $X^{\prime M}={\cal F}^{M}{}_{N}X^{N}$, 
which in general is different from the actual $X^{\prime}$.}  Raising the $N$ index we have
\be
\delta_M{}^N \ = \ {\cal F}_M{}^R \,{\cal F}^N{}_R \ = \
 ({\cal F\, F}^t\, )_M^{~~N} \,, 
\ee
and therefore we must check that 
\be
{\cal F \,F}^t \ = \ {\bf 1}  \,. 
\ee
We thus calculate with (\ref{calFdefcalc})   
\be
{\cal FF}^t \ = \  {1\over 2} 
  \Bigl(  {1\over 1-a}  
  \, (1-a^t)
  + (1-a^t) {1\over 1-a} \Bigr)
   \,{1\over 2} 
  \Bigl( (1-a) {1\over 1-a^t}   \,  
  + {1\over 1-a^t} (1-a)   \Bigr)\;. 
\ee 
The cross terms give multiples of the unit matrix, but the other two terms
are more complicated, 
\be
\label{dflkdf}
{\cal FF}^t \ = \  {1\over 2}  + {1\over 4} \Bigl( 
{1\over 1-a}  
  \, (1-a^t)(1-a) {1\over 1-a^t}  +  (1-a^t) {1\over 1-a} {1\over 1-a^t} (1-a) 
  \Bigr) \;. 
\ee 
We note that if the order of the second and third factors in the first term was opposite we would have a simple product.  The same holds for the first and second factors in the second term. The computation is thus helped by the use of the following commutators:
\be
\bigl[ 1- a^t , 1-a \bigr] \ = \ - aa^t  \,, ~~~
\Bigl[ 1-a^t , {1\over 1-a} \Bigr] \ = \ {1\over 1-a}  \, aa^t \,.
\ee
With these (\ref{dflkdf}) becomes
\be
\label{dflkdf-vm}
{\cal FF}^t \ = \  1  + {1\over 4} \Bigl( -
{1\over 1-a} \, aa^t   
 \, {1\over 1-a^t}  + {1\over 1-a} \, aa^t {1\over 1-a^t}  (1-a) 
  \Bigr) \;. 
\ee 
The terms in parenthesis cancel: in the second one we can 
bring the $a^t$ in $aa^t$ to the right, where it kills $a$.  We thus
proved that ${\cal F\,F}^t = {\bf 1}$.  This implies the desired
gauge invariance of $\eta$ or, equivalently, its independence
of the chosen coordinate system. Moreover, it proves  
that ${\cal F}_M{}^N$ is an $O(D,D)$ matrix.

It is also straightforward to verify that, as expected, ${\cal F}$ and 
${\cal F}^t$ are also inverses of each other in the other direction:
 \be
  {\cal F}^t {\cal F} \ = \ 1\;.
 \ee
Indeed, this time we get
\be
\label{dflkdfvmpss}
{\cal F}^t{\cal F} \ = \  {1\over 2}  + {1\over 4} \Bigl( 
 (1-a) {1\over 1-a^t} {1\over 1-a} (1-a^t) 
 + {1\over 1-a^t}  
  \, (1-a)(1-a^t) {1\over 1-a}   
  \Bigr) \,.
\ee 
The simplest way to evaluate the left-over terms is to expand
using $a^t a=0$.  Each of the two summands gives in fact
simple expressions:
\be
\label{dflkdfbtflvmpss}
{\cal F}^t{\cal F} \ = \  {1\over 2}  + {1\over 4} \Bigl( 
 (1- a a^t)   \ + \  (1 + aa^t) 
  \Bigr) \ = \ {\bf 1} \,.
  \ee

The coordinate transformation for a generalized tensor with an upper
index is obtained from  (\ref{finiteF})
by raising the index:  
 \be
  A^{\prime M}(X^{\prime}) \ = \ {\cal F}^{M}{}_{N} A^{N}(X)\; .
 \ee
Of course, the indices on ${\cal F}$ are 
 raised and lowered with $\eta$, so that (\ref{calFdef}) gives
 \be
  {\cal F}^{M}{}_{N} \ = \ \frac{1}{2}\Big(\frac{\partial X_{P}}{\partial X^{\prime}_{M}}\,\frac{\partial X^{\prime P}}{\partial X^{N}}+
  \frac{\partial X^{\prime M}}{\partial X^{P}}\,\frac{\partial X_{N}}{\partial X^{\prime}_{P}} \Big)\;. 
 \ee
Consistent with the invariance of $\eta$, it follows that the contraction of upper and lower indices gives a tensor of lower 
rank, e.g., 
 \be
  A^{\prime M}B^{\prime}_{M} \ = \ {\cal F}^{M}{}_{N}A^{N}{\cal F}_{M}{}^{K}B_{K} \ = \ 
  A^{N}({\cal F}^t{\cal F})_{N}{}^{K}B_{K} \ = \ A^{N}\delta_{N}{}^{K} B_{K} \ = \ A^{N}B_{N}\;.
 \ee

Let us comment on inverse transformations.  If we perform a coordinate
transformation $X \to X'$ followed by $X' \to X$ the result should
be no coordinate transformation.  In the notation of (\ref{inmatrixF}) we 
should have 
\be
\label{testinv}
 {\cal F}_M{}^N (X , X') \,  {\cal F}_N{}^P (X' , X)  \ = \ \delta_M{}^P \,.
\ee       
As we would expect, this is closely related to the $O(D,D)$ properties
of ${\cal F}$ noted above.  We see from (\ref{inmatrixFvm})
      \be  
 \label{inmatrixFvm99}  
 \begin{split}
 {\cal F} (X , X') \  = & \  \   {1\over 2} \,
  \Bigl( \,  \frac{\partial X'}{\partial X}\,
  \Bigl(\frac{\partial X}{\partial X'}\Bigr)^t
  +    \Bigl(\frac{\partial X}{\partial X'}\Bigr)^t\,
  \frac{\partial X'}{\partial X}\, \Bigr)  \\
   = & \  \   {1\over 2} \,
  \Bigl( \,  \frac{\partial X}{\partial X'}\,
  \Bigl(\frac{\partial X^{\prime}}{\partial X}\Bigr)^t
  +    \Bigl(\frac{\partial X'}{\partial X}\Bigr)^t\,
  \frac{\partial X}{\partial X'}\, \Bigr)^t \ = \  {\cal F} (X', X)^t  \,. 
\end{split}
 \ee
With indices, we write
\be
{\cal F}_M{}^N (X , X')\ = \   {\cal F}^N{}_M (X' , X)\,. 
\ee      
Back on the left-hand side of (\ref{testinv}) we have
\be
\label{testinvbtflvm}
 {\cal F}^N{}_M (X' , X) \,  {\cal F}_N{}^P (X' , X)  \ = \ 
 ({\cal F}^t  {\cal F})_M{}^{P}   \ = \ \delta_M{}^P \,.
\ee       
This confirms that the postulated transformation is
consistent with the independent definition of the inverse.

Our computations used at various points the strong constraint. This 
constraint implies unusual relations.  For example we have
found that
\be
-1 +  \frac{\partial X}{\partial X'}  +  \Bigl( \frac{\partial X^{\prime }}{\partial X}\Bigl)^t  \ = \   \Bigl( \frac{\partial X^{\prime }}{\partial X}\Bigl)^t 
   \frac{\partial X}{\partial X'} \;, 
\ee
which is readily checked 
using (\ref{px'px})  and (\ref{pxpx'}).  This relation allows us to 
write  ${\cal F}$ differently, but not in any simpler way.
Using the above and (\ref{inmatrixFin}) we have, for example,  
\be
{\cal F} \ = \ 
-1 +  \frac{\partial X}{\partial X'}  +  \Bigl( \frac{\partial X^{\prime }}{\partial X}\Bigl)^t  + {1\over 2}  \Bigl[  \frac{\partial X}{\partial X'} \,, \, 
   \Bigl( \frac{\partial X^{\prime }}{\partial X}\Bigl)^t \Bigr]\,   .   
\ee  
Using relations like this   
we have experimented with various other candidate expressions 
for ${\cal F}$, but have 
not found 
an equally natural expression 
that passes all consistency requirements.

\section{Special  gauge transformations and $O(D,D)$ 
}\label{spegautraandoddtra}

The purpose of this section is two-fold.  We first show, in subsection~\ref{gencooandbfi},  
how the 
standard, finite coordinate transformations of the 
non-doubled fields arise from the finite transformations
generated by ${\cal F}$ in the doubled theory.  
In subsection~\ref{therelbetoddandgausym} we 
discuss  
to what extent finite $O(D,D)$
transformations are contained in the gauge group. 
Viewing the $O(D,D)$
rotation of coordinates directly as a generalized coordinate transformation
leads to a puzzling result: the gauge transformed field and the $O(D,D)$ transformed field differ by one power of the $O(D,D)$ rotation. 
Resolving this paradox 
we find that only the 
geometric subgroup $GL(D,\mathbb{R})\ltimes \mathbb{R}^{\frac{1}{2}D(D-1)}$ can always
be realized as special coordinate transformations, but that in the context of a  
reduction on the torus $T^d$, 
the full $O(d,d)$ subgroup of $O(D,D)$  is part of the gauge group.

\subsection{General coordinate and $b$-field  
 gauge transformations} 
\label{gencooandbfi}
We will now show that the postulated finite coordinate transformations in double field theory reduce for special 
cases to the standard finite gauge transformations, namely general coordinate 
transformations and $b$-field  
 gauge transformations. 
It turns out that these transformations, c.f.~(\ref{specialcoord1}), (\ref{specialcoord2})
and (\ref{bgauge}) below, are special transformations $X\rightarrow X^{\prime}$ for which the two 
terms in (\ref{delAM}) are actually equal so that ${\cal F}$ simplifies to one term, 
as in (\ref{aatzero}), 
  \be\label{coordconstr}
  \frac{\partial X^{\prime}_{ M}}{\partial X_{P}}\,
  \frac{\partial X^{N}}{\partial X^{\prime P}} \ = \ 
  \frac{\partial X^{P}}{\partial X^{\prime M}}\,
  \frac{\partial X^{\prime}_{ P}}{\partial X_{N}}\,\qquad \Rightarrow \qquad
  {\cal F}_{M}{}^{N} \ = \  \frac{\partial X^{\prime}_{ M}}{\partial X_{P}}\,
  \frac{\partial X^{N}}{\partial X^{\prime P}} \ = \ 
  \frac{\partial X^{P}}{\partial X^{\prime M}}\,
  \frac{\partial X^{\prime}_{ P}}{\partial X_{N}} \;.  
 \ee
We recall from (\ref{aatzero}) that this holds if $aa^t=0$, which by (\ref{px'px}) means 
 \be
  (aa^t)_{MN} \ = \ a_{M}{}^{P}a_{NP} \ = \ \partial_{M}\zeta^{P}\,\partial_{N}\zeta_{P}\;.
 \ee
If we have either $\zeta^i=0$ or $\tilde{\zeta}_i=0$ the
$O(D,D)$ invariant sum  over 
$P$ vanishes and  
(\ref{coordconstr}) holds. 
This will apply below, since we will consider
general coordinate 
and $b$-field gauge transformations separately.

We start with a vector $A_{M}(x)$
independent of $\tilde{x}$ and a coordinate transformation
 \be\label{specialcoord1}
  x^{i}\;\rightarrow\; x^{i\prime} \  = \ x^{i\prime}(x)\;, \qquad \tilde{x}_{i}^{\prime} \ = \ \tilde{x}_{i}\;.
 \ee          
Since this transformation leaves $\tilde{x}_i$ invariant, the corresponding parameter $\tilde{\zeta}_i$ is zero, 
and thus we can apply (\ref{coordconstr}). 
Specializing (\ref{finiteF}) to $A_{i}$ and using the second form of ${\cal F}$ in (\ref{coordconstr})  we get 
 \be
  A_{i}^{\prime}(x^{\prime}) \ = \ \frac{\partial X^{P}}{\partial x^{i\prime}}\,\frac{\partial X^{\prime}_{P}}{\partial X_{N}}\,A_{N}(x)
  \ = \ \frac{\partial x^{p}}{\partial x^{i\prime}}\,\frac{\partial \tilde{x}^{\prime}_{p}}{\partial \tilde{x}_{n}}\,A_{n}(x)
  \ = \ \frac{\partial x^{p}}{\partial x^{i\prime}}\,\delta_{p}^{n}\,A_{n}(x) \ = \ 
  \frac{\partial x^{p}}{\partial x^{i\prime}}\,A_{p}(x)\;, 
 \ee
which is precisely the standard general coordinate 
transformation of a co-vector. Specializing (\ref{finiteF}) to $A^{i}$ we get
 \be
  A^{i\prime}(x^{\prime}) \ = \ \frac{\partial X^{P}}{\partial \tilde{x}^{\prime}_i}\, \frac{\partial X_{P}^{\prime}}{\partial X_{N}}\,A_{N}(x)
  \ = \ \frac{\partial \tilde{x}_{p}}{\partial \tilde{x}^{\prime}_i}\, \frac{\partial x^{p \prime}}{\partial x^{n}}\,A^{n}(x)
  \ = \ \delta_{p}^{i}\,\frac{\partial x^{p \prime}}{\partial x^{n}}\,A^{n}(x) \ = \ \frac{\partial x^{i \prime}}{\partial x^{n}}\,A^{n}(x)\;, 
 \ee              
which is the general coordinate 
transformation of a vector.

If we consider now a field depending only on $\tilde{x}$ and a transformation 
 \be\label{specialcoord2}
   \tilde{x}_{i}\;\rightarrow\; \tilde{x}_{i}^{\prime} \  = \ \tilde{x}_{i}^{\prime}(\tilde{x})\;, \qquad {x}^{i \prime} \ = \ x^{i}\;, 
 \ee   
that transforms only the $\tilde{x}$
we have $\zeta^i=0$ and so we can again apply (\ref{coordconstr}). We get by a completely analogous computation 
 \be
  A_{i}^{\prime}(\tilde{x}^{\prime}) \ = \ \frac{\partial \tilde{x}^{\prime}_{i}}{\partial \tilde{x}_{n}}\,A_{n}(\tilde{x})\;, \qquad
  A^{i\prime}(\tilde{x}^{\prime}) \ = \ \frac{\partial \tilde{x}_p}{\partial \tilde{x}_i^{\prime}}\,A^{p}(\tilde{x})\;. 
 \ee
Therefore, they transform conventionally, where we recall that for dual coordinate transformations  
the notion of covariant and contravariant indices is interchanged.

\medskip

Let us now consider 
$b$-field gauge transformations,
which should follow from 
 \be\label{bgauge}
  \tilde{x}_i^{\prime} \ = \ \tilde{x}_i-\tilde{\zeta}_i(x)\;, \qquad x^{i\prime} \ = \ x^i\;.
 \ee
As $\tilde{\zeta}_i$ depends on $x$ this transformation mixes $x$ and $\tilde{x}$, but 
still  
satisfies 
condition (\ref{coordconstr}) 
since  
$\zeta^i=0$. We first compute 
 \be
  \frac{\partial X^{\prime M}}{\partial X^{N}} \ = \      
   \begin{pmatrix}   \frac{\partial \tilde{x}_i^{\prime}}{\partial \tilde{x}_j} & \frac{\partial x^{i \prime}}{\partial \tilde{x}_j}\\[0.5ex]
  \frac{\partial \tilde{x}_i^{\prime}}{\partial x^j} & \frac{\partial x^{i \prime}}{\partial x^j}\end{pmatrix}
  \ = \ \begin{pmatrix}   \delta_i{}^{j} & 0 \\[0.5ex]
  -\partial_j\tilde{\zeta}_i & \delta_j{}^{i} \end{pmatrix}\;, 
 \ee
and the inverse 
   \be
  \frac{\partial X^{M}}{\partial X^{\prime N}} \ = \      
   \begin{pmatrix}   \frac{\partial \tilde{x}_i}{\partial \tilde{x}_j^{\prime}} &  \frac{\partial x^{i }}{\partial \tilde{x}_j^{\prime}} \\[0.5ex]
  \frac{\partial \tilde{x}_i}{\partial x^{j\prime}}  & \frac{\partial x^{i}}{\partial x^{j\prime}}\end{pmatrix}
  \ = \ \begin{pmatrix}   \delta_i{}^{j} & 0 \\[0.5ex]
   \partial_j\tilde{\zeta}_i & \delta_j{}^{i} \end{pmatrix}\;. 
 \ee 
We will now show that (\ref{bgauge}) indeed leads to the expected $b$-field gauge transformations. 
We apply a finite gauge transformation to the generalized metric    
 \be\label{firstH}
  {\cal H}_{MN} \ = \  \begin{pmatrix}    {\cal H}^{ij} & {\cal H}^{i}{}_{j}\\[0.5ex]
  {\cal H}_{i}{}^{j} & {\cal H}_{ij}  \end{pmatrix}
  \ = \ \begin{pmatrix}    g^{ij} & -g^{ik}b_{kj}\\[0.5ex]
  b_{ik}g^{kj} & g_{ij}-b_{ik}g^{kl}b_{lj}\end{pmatrix}\;,
 \ee
which reads 
 \be
  {\cal H}_{MN}^{\prime}(X^{\prime}) \ = \ \frac{\partial X^P}{\partial X^{\prime M}}\,\frac{\partial X^{\prime}_P}{\partial X_K}\,
  \frac{\partial X^Q}{\partial X^{\prime N}}\,\frac{\partial X^{\prime}_Q}{\partial X_L}\,{\cal H}_{KL}(X)\;. 
 \ee 
Specializing  
to the component ${\cal H}^{ij}$, we get 
 \be
  {\cal H}'^{\,ij}\ = \  
  \frac{\partial X^P}{\partial \tilde{x}^{\prime}_i}\,
  \frac{\partial X^{\prime}_P}{\partial X_K}\,
  \frac{\partial X^Q}{\partial \tilde x'_{j}}\,
  \frac{\partial X^{\prime}_Q}{\partial X_L}
  \,{\cal H}_{KL}\;, 
 \ee 
and we assume that ${\cal H}$ depends initially only on $x$ so that by (\ref{bgauge}) ${\cal H}^{\prime}$ has the same 
coordinate dependence, which we suppress.  Inserting the non-vanishing derivatives we get
\be
 {\cal H}'^{\,ij}\ = \  
  \frac{\partial \tilde x_p}{\partial \tilde{x}^{\prime}_i}\,
  \frac{\partial x'^p}{\partial x^k}\,
  \frac{\partial \tilde x_q}{\partial \tilde x'_{j}}\,
  \frac{\partial x'^q}{\partial x^l}
  \,{\cal H}^{kl}\ = \ \delta^i_p\,\delta^p_k\, \delta^j_q \,\delta^q_l \, 
  {\cal H}^{kl} \ = \ {\cal H}^{ij} \;, 
\ee
and comparing with (\ref{firstH}) we deduce that
\be
\label{gb-ftran}
g^{ij\,\prime} \ = \ g^{ij}\;.
\ee
Thus, as expected, the metric is invariant under $b$-field gauge transformations. 
Specializing now  
 to the component ${\cal H}^{i}{}_{j}$ and inserting the non-vanishing
 derivatives  we get 
 \be
 \begin{split}
  {\cal H}^{i}{}_{j}{}^{\prime} \ & = \  \frac{\partial X^P}{\partial \tilde{x}^{\prime}_i}\,\frac{\partial X^{\prime}_P}{\partial X_K}\,
  \frac{\partial X^Q}{\partial x^{j \prime}}\,\frac{\partial X^{\prime}_Q}{\partial X_L}\,{\cal H}_{KL} \\ 
\ &= \  \frac{\partial \tilde{x}_p}{\partial \tilde{x}^{\prime}_i}\,\frac{\partial x^{\prime p}}{\partial x^k}\,
  \frac{\partial x^q}{\partial x^{j \prime}}\,\frac{\partial \tilde{x}^{\prime}_q}{\partial \tilde{x}_l}\,{\cal H}^{k}{}_{l}
  + \frac{\partial \tilde{x}_p}{\partial \tilde{x}^{\prime}_i}\,\frac{\partial x^{\prime p}}{\partial x^k}\,
  \frac{\partial x^q}{\partial x^{j \prime}}\,\frac{\partial \tilde{x}^{\prime}_q}{\partial x^l}\,{\cal H}^{kl}
  +\frac{\partial \tilde{x}_p}{\partial \tilde{x}^{\prime}_i}\,\frac{\partial x^{\prime p}}{\partial x^k}\,
  \frac{\partial \tilde{x}_q}{\partial x^{j \prime}}\,\frac{\partial x^{q \prime}}{\partial x^l}\,{\cal H}^{kl} \\
  \ &= \ \delta_p{}^i\, \delta_k{}^p\,\delta_j{}^q\,\delta_q{}^l\,{\cal H}^{k}{}_{l}
  +\delta_p{}^i\, \delta_k{}^p\,\delta_j{}^q\,(-\partial_l\tilde{\zeta}_q){\cal H}^{kl}
  +\delta_p{}^i\, \delta_k{}^p\,(\partial_j\tilde{\zeta}_q)\delta_l{}^q\,{\cal H}^{kl} \\
  \ &= \ {\cal H}^{i}{}_{j}-\partial_l\tilde{\zeta}_j\, {\cal H}^{il}+\partial_j\tilde{\zeta}_l\,{\cal H}^{il}\, .   
 \end{split}
 \ee  
Making use of 
(\ref{firstH}) we then find that   
 \be
  -g^{ik\,\prime}\,b_{kj}^{\prime} \ = \ -g^{ik}b_{kj}-g^{ik}(\partial_k\tilde{\zeta}_j-\partial_j\tilde{\zeta}_k)\;.
 \ee
From this and (\ref{gb-ftran}) 
we infer that
 \be\label{bfiedlgauge}
b_{ij}^{\prime} \ = \ b_{ij}+ 
 \partial_i\tilde{\zeta}_j-\partial_j\tilde{\zeta}_i\,, 
 \ee
showing that 
the generalized coordinate transformations reproduce precisely the finite  
$b$-field gauge transformations.

\medskip

\subsection{The relation between 
$O(D,D)$ and gauge symmetries} \label{therelbetoddandgausym}
We ask now to what extent $O(D,D)$ transformations are 
generalized coordinate transformations. 
Consider the finite $O(D,D)$
transformation   
 \be\label{ODDansatz}
  X^{\prime M} \ = \ h^{M}{}_{N}X^{N}\,,   ~~~\hbox{or}~~~\quad  X' \ = \ h \, X
    \;,  
 \ee
which, by definition, 
 acts on a vector field as 
  \be\label{trueODD}
  A_{M}^{\prime}(X^{\prime})   
  \ = \ A_{N}(X) \big(h^{-1}\big)^{N}{}_{M}\, ~~~\hbox{or} ~~~\quad  
    A' (X'= hX) \ = \  A(X)\,  h^{-1} \,. 
 \ee

As a first naive attempt let us view (\ref{ODDansatz}) as a generalized 
coordinate transformation and compute its action on a vector $A_{M}(X)$. 
The derivatives 
are 
 \be
  \frac{\partial X^{\prime M}}{\partial X^{N}} \ = \ h^{M}{}_{N}\;, \qquad 
  \frac{\partial X^{M}}{\partial X^{\prime N}} \ = \ (h^{-1} )^{M}{}_{N}\;, 
 \ee
or in matrix notation 
 \be
  \frac{\partial X^{\prime}}{\partial X} \ = \ h^t \;, \qquad
 \frac{\partial X}{\partial X^{\prime}} \ = \ (h^{-1})^t\;.
 \ee 
We can then use (\ref{inmatrixF}) to write the gauge transformation, 
including the $O(D,D)$ metrics   
 that are implicit in (\ref{calFdef})  in the $PP$ contractions and the coordinates with lowered indices:   
 \be
  A_{M}^{\prime}(X^{\prime}) \ = \ \frac{1}{2}\Big( (h^{-1})^t\eta^{-1}h\eta+\eta^{-1}h\eta (h^{-1})^t\Big)_{M}{}^{N}A_{N}(X)\;. 
 \ee
We have $h\eta h^{t} = \eta$, from 
which we conclude for the first term 
 \be
  (h^{-1})^t\eta^{-1}h\eta \ = \  (h^{-1})^t\eta^{-1}\eta (h^t)^{-1} \ = \  \big[ (h^{-1})^t\big]^2 \;, 
 \ee
and for the second 
 \be
  \eta^{-1}h\eta (h^{-1})^t \ = \ \eta^{-1}\eta(h^t)^{-1}(h^{-1})^t  \ = \  \big[ (h^{-1})^t\big]^2 \;. 
 \ee 
Thus, the transformation rule is 
 \be\label{squarerule-}
  A_{M}^{\prime}(X^{\prime}) \ = \ \Big[\big( (h^{-1})^t\big)^2\Big]_{M}{}^{N}A_{N}(X) \ = \ \,A_{N}(X)
  \big[\big(h^{-1}\big)^2\big]^{N}{}_{M}\;. 
 \ee
 In index-free notation,
 \be\label{squarerule}
  A^{\prime}(X^{\prime}= hX) \ = \ \,A(X)
  (h^{-1})^2\;. 
 \ee
Comparing with (\ref{trueODD}) we infer that the gauge symmetry gives the square of the matrix we want! This is the finite 
version of 
the same phenomenon 
encountered at the infinitesimal level in \cite{Hohm:2010jy}.
There we saw that the infinitesimal version of the naive ansatz (\ref{ODDansatz}) leads to 
a relative factor of two between the transport term and the rest.      

The reason that the above does not indicate an inconsistency is that,
viewed as a general coordinate transformation,  the ansatz (\ref{ODDansatz}) is not allowed in general   
by the strong constraint.    
We will use (\ref{squarerule}) as a guide to modify the generalized
coordinate transformation associated to the duality transformation (\ref{ODDansatz}).
While the coordinate transformation will differ from the duality
transformation in the way coordinates are rotated, the field 
 transformations can be made to agree, under conditions 
 to be explained below.

Consider first the geometric subgroup $GL(D,\mathbb{R})\ltimes \mathbb{R}^{\frac{1}{2}D(D-1)}$ of $O(D,D)$, 
whose elements do 
not mix the $x$ and $\tilde{x}$ coordinates.
This subgroup, we claim,
can be realized as  
(generalized) coordinate transformations. 
To prove this claim, we work in a frame 
 in which the fields do not depend on~$\tilde x$.
Consider the dualities defined by a constant
 $\Lambda\in GL(D,\mathbb{R})$  
embedded in $O(D,D)$ as  $\Lambda\rightarrow h(\Lambda)$, 
with 
 \be
 \label{xu}
  (h^{-1})^{N}{}_{M}(\Lambda) \ = \   \begin{pmatrix}    (h^{-1})_j{}^{i} & (h^{-1})_{ji}\\[0.5ex]
  (h^{-1})^{ji} & (h^{-1})^{j}{}_{i}  \end{pmatrix} \ = \ \begin{pmatrix}   \Lambda^{i}{}_j & 0\\[0.5ex]
  0 & (\Lambda^{-1})^{j}{}_{i}  \end{pmatrix}\;. 
 \ee 
The corresponding $O(D,D)$ transformation (\ref{trueODD}) on a vector $A_{M}=(A^{i},A_{i})$ then gives 
 \be
 \label{xxyz}
  A_i^{\prime}(x^{\prime}) \ = \ (\Lambda^{-1})^{j}{}_{i}\,A_{j}(x)\;, \qquad
  A^{i\prime}(x^{\prime}) \ = \ \Lambda^{i}{}_j\,A^{j}(x)\;, 
 \ee 
where only the transformation of $x$ is relevant in the argument of the fields.  The associated  generalized coordinate transformation
is
 \be\label{GLDcoord}
  x'^{i} \ = \ \Lambda^{i}{}_{j}\,x^{j}\;, \qquad \tilde{x}_{i}' \ = \ \tilde{x}_{i}\;, \qquad \Lambda\in GL(D,\mathbb{R})\;.
 \ee
As anticipated above, this
is not the coordinate rotation induced by 
 $GL(D,\mathbb{R})\subset O(D,D)$, which would
 also transform  
 $\tilde{x}$ (in the dual representation according to (\ref{xu})). 
Equation (\ref{GLDcoord}) is a special case of (\ref{specialcoord1}),
so we can use the results of that subsection to find that
this coordinate transformation yields    
 \be
  A_i^{\prime}(x^{\prime}) \ = \ (\Lambda^{-1})^{j}{}_{i}\,A_{j}(x)\;, \qquad
  A^{i\prime}(x^{\prime}) \ = \ \Lambda^{i}{}_j\,A^{j}(x)\;,  
 \ee 
 resulting in complete agreement with
(\ref{xxyz}). 

Finally, consider now the constant shift transformations in 
the duality subgroup
$\mathbb{R}^{\frac{1}{2}D(D-1)}$ of  $O(D,D)$.
These, with constant parameter $e_{ij}=-e_{ji}$,  are given by 
 \be
 \label{yu}
   (h^{-1})^{N}{}_{M}(e)  
\ = \ \begin{pmatrix}   \delta^{i}{}_j & -e_{ij}\\[0.5ex]
  0 & \delta^{j}{}_{i}  \end{pmatrix}\;.  
 \ee 
It is easy to check that this acts on the generalized metric by $b_{ij}\rightarrow b_{ij}+e_{ij}$.   We claim that the associated  
 generalized coordinate transformations  are
 \be\label{bgauge2}
  \tilde{x}_i^{\prime} \ = \ \tilde{x}_i+\frac{1}{2}e_{ij}x^{j}\;, \qquad x^{i\prime} \ = \ x^i\;. 
 \ee
 Again, this differs (by a factor of two) from the coordinate
 transformations suggested by the dualities (\ref{yu}).  
Equations~(\ref{bgauge2})  are a special case of (\ref{bgauge}), applicable for fields that depend only on $x$, and also result in 
$b_{ij} \to b_{ij} + e_{ij}$.  
Summarizing, the full geometric subgroup is part of the gauge group.

Let us now turn to the remaining transformations that complete the geometric subgroup 
to the full T-duality group $O(D,D)$. 
Instead of (\ref{ODDansatz}) we  consider the 
generalized coordinate transformation  
 \be\label{rootrot}
  X^{\prime M} \ = \ \big(\sqrt{h}\,\big)^{M}{}_{N}X^{N}\;. 
 \ee
The square root of the group element always exists and is itself a group element for the component 
connected to the identity: we may simply insert a factor of $\tfrac{1}{2}$ in the exponential representation 
of $h$ in order to construct $\sqrt{h}$. Since $\sqrt{h}$ is an $O(D,D)$ element the above computation 
leading to (\ref{squarerule}) proceeds in exactly the same way, but now we obtain 
 \be\label{fakeODD-}
  A_{M}^{\prime}(X^{\prime}) \ = \  \big[\big((\sqrt{h})^{-1}\big)^{2}\big]^{N}{}_{M}\,A_{N}(X) 
  \ = \  A_{N}(X)\big(h^{-1}\big)^{N}{}_{M}\,\;.
 \ee
 More schematically, and without indices, we write
 \be\label{fakeODD}
  A^{\prime}(X^{\prime}= \sqrt{h}\,X) \ = \   A(X)
  \, h^{-1}\,\;.
 \ee
The right-hand side is as required by the $O(D,D)$ transformation (\ref{trueODD}), 
but the left-hand side is not, because 
 $X^{\prime}=\sqrt{h}\,X$ rather than $X'= hX$.
We conclude that in general the full $O(D,D)$ cannot be seen as part of the gauge group. However, 
for the special case that 
the fields depend only on a subset of half of the coordinates that are allowed by the strong constraint the situation changes. 
In this case we can consider $O(D,D)$ transformations that act only on those coordinates on which the fields do 
not depend. We then have $A^{\prime}(X^{\prime})=A^{\prime}(X)$ and the two formulas (\ref{trueODD}) and (\ref{fakeODD}) coincide.   
We use this approach now to 
see that the remaining $O(D,D)$ transformations 
can be realized as coordinate transformations,   
consistent with the strong constraint.
We already have the group elements  (\ref{xu}) and (\ref{yu}). To generate
the full $O(D,D)$ we are missing the elements
 \be\label{zu}
  h^{M}{}_{N}(f) \ = \ \begin{pmatrix}   \delta_{i}{}^j & 0\\[0.5ex]
  f^{ij} & \delta^{i}{}_{j}  \end{pmatrix} \qquad \Rightarrow \qquad 
  \big(\sqrt{h}\,\big)^{M}{}_{N}(f) \ = \ \begin{pmatrix}   \delta_{i}{}^j & 0\\[0.5ex]
  \frac{1}{2} f^{ij} & \delta^{i}{}_{j}  \end{pmatrix}\;. 
 \ee
The coordinate transformation (\ref{rootrot}) then reads 
\be
  X^{\prime M} \ = \ \big(\sqrt{h}\,\big)^{M}{}_{N}X^{N}\qquad \Rightarrow \qquad
  \tilde{x}_i^{\prime} \ = \ \tilde{x}_i\;, \quad x^{i\prime} \ = \ x^i +\frac{1}{2}f^{ij}\tilde{x}_{j}\;. 
 \ee
The last equation implies $\tilde{\zeta}_i=0$, $\zeta^i =  -\frac{1}{2}f^{ij}\tilde{x}_{j}$, 
and thus the gauge parameters depend only on the $\tilde{x}_i$ on which the above transformation acts. 
As discussed above, the fields 
are now assumed to be independent of the dual 
$x^i$ coordinates, so  the strong constraint is satisfied.
Therefore we have shown that these particular $O(D,D)$ transformations are special gauge transformations. 
In other words, in the case of  a torus reduction, where the fields are independent of 
$d<D$ (internal) coordinates, we can view the full $O(d,d)$ subgroup as part of 
the gauge group. This analysis completes our previous analysis in \cite{Hohm:2010jy} for the case of finite  gauge transformations.

\section{Exponentiation of generalized Lie derivatives} 
In this section we compare the postulated 
formula (\ref{finiteF})  
for generalized coordinate transformations 
with an 
 alternative definition of finite 
transformations  
as the result of exponentiation of generalized Lie derivatives
$\widehat{\cal L}_\xi$, with parameter $\xi$. We determine
how the parameter $\xi$ 
enters into the generalized coordinate transformation $X\rightarrow X^{\prime}$
to quartic order in $\xi$ and verify the resulting  
equivalence of the two forms of finite  transformations to that
order.

\subsection{General coordinate transformations}
We start by writing a finite coordinate transformation in terms of 
a parameter $\xi^{M}(X)$ that generates this transformation as follows 
\be\label{newXprime}
  X^{M}\; \rightarrow\;   X^{\prime M} 
  \ = \ e^{-\xi^{P}(X)\partial_P} X^{M}\;. 
 \ee  
In this right-hand side the exponential is meant to be expanded
in a power series and the differential operator $\xi^M\partial_M$, written sometimes
 as $\xi$, acts to the right on a function to give a function. 
We can also rewrite (\ref{newXprime}) as an operator equation as follows
   \be\label{Xprimenew}
    X^{\prime} \ = \ e^{-\xi}\,X\,e^{\xi}\; .
   \ee
This can be verified with the familiar relation
$e^{A}\,B\, e^{-A} =  B+[A,B] +\frac{1}{2}[ A,[A,B]]+\cdots$, 
recalling that for any function $f(X)$ we have
$[\xi , f(X)]  =  \xi^M \partial_M f$. 
Equation (\ref{Xprimenew}) is to be interpreted as an operator equation, 
in which the left-hand side is a function that is 
viewed  as an operator acting via multiplication.  

  The parameter $\xi$ can be  
  related to $\zeta$   
  defined in (\ref{xi}),  
  $\zeta^{M}  =  \xi^{M}-\frac{1}{2}\xi^{P}\partial_{P}\xi^{M}+{\cal O}(\xi^{3})$, but this 
  will not be required in the discussion
  that follows.  The $\xi$  parameterization of the coordinate
  change will be used henceforth unless noted otherwise.
  We could write $X'^M_\xi$ to denote the $\xi$ dependence but
  we will not do so unless it is required to distinguish it from other
  possible definitions of $X'$.
We write the above diffeomorphism more compactly as
\be
\label{diffmc}
X' \ = \ e^{-\xi} X\ = \ \Bigl( 1
- \xi + {1\over 2} \xi^2 - \ldots\Bigr)X \,, \qquad  \xi \ \equiv \ \xi^M\partial_M \,.
\ee
 Taking derivatives of $X'$
with respect to $X$ is not 
complicated and one quickly
finds that
 \be
 \label{primepartialnormal-vcm}
  \frac{\partial X^{\prime }}{\partial X} \ = \ {\bf 1}-a
  +\frac{1}{2}(\xi+a)a
  -\frac{1}{3!}(\xi+a)^2a+\frac{1}{4!}(\xi+a)^3 a+\cdots\;,
  ~~a_P{}^Q \equiv  \partial_P \xi^Q \,. 
 \ee
In here the $\xi$ operator acts on everything that stands to its right. 
 For example, $\xi a^2 =  (\xi a) a + a (\xi a)$.  The 
above right-hand side is a (matrix) function, not a
(matrix) differential  
operator. Letting the $\xi$ act we have
\be
 \label{primepartialnormal-vmbv} 
  \frac{\partial X^{\prime }}{\partial X} \ =  \ {\bf 1}-a
  \ + \ \frac{1}{2}\bigl( \xi a  + \,a^2\bigr)  \ 
  -\ \frac{1}{6}\bigl( \xi^2 a  
  + (\xi a)a   + 2a\,\xi a  + a^3\,\bigr) \ + \  {\cal O} (\xi^4) \,. \ee
Equation (\ref{primepartialnormal-vcm}) can be written as
 \be
 \label{ilvyvm}
  \frac{\partial X^{\prime }}{\partial X} \ = \  \Bigl( e^{- (\xi + a)}  \,{\bf 1} \Bigr) \,,
 \ee
where the full expansion of the exponential acts on the constant
matrix ${\bf 1}$.   Since $(\xi+ a) {\bf 1} =  a$, one sees immediately that the evaluation of (\ref{ilvyvm}) yields
(\ref{primepartialnormal-vcm}).  
 We now claim that we can
simply write 
\be
\label{vmlvlvm}
 \frac{\partial X^{\prime }}{\partial X} \ = \  
  e^{- (\xi + a)}  e^\xi  \,.
\ee
Here the right-hand side may seem to be a differential
operator but it is in fact a function, the function  
given in~(\ref{ilvyvm}). 
To prove this define $h(t)$ by 
\be
h(t) \ \equiv \  e^{-t (\xi + a)}  e^{t\xi}\,\,. 
\ee
Taking a derivative of $h$ with respect to $t$ we get
$h' (t) =   e^{-t (\xi + a)}  (-a)  e^{t\xi}$,
and note that the object in between the exponentials is
a function, not a differential 
 operator.  We can write this~as
\be
h' (t) \ = \  e^{-t (\xi + a)}  \bigl(-(\xi +a) {\bf 1} \bigr)   e^{t\xi}\,. 
\ee
Then passing from the $n$-th derivative to the next
goes as follows
\be
h^{(n)} (t) = e^{-t (\xi + a)}  g_n\, e^{t\xi }\quad \to \quad
h^{(n+1)} (t) = e^{-t (\xi + a)} \bigl( -(\xi + a) g_n\bigr) \, e^{t\xi}\,.
\ee
We note that in the expression for $h^{(n+1)}$ the operator $\xi$
acts only on $g_{n}$, because the term where it acts on $e^{t\xi}$
cancels against the derivative of $e^{t\xi}$. 
If $g_n$ is a function, the object in between the exponentials
in $h^{(n+1)}$ is also a function.  The result now follows because
the above establishes that 
$h^{(n)} (t=0) =  (-1)^n  (\xi + a)^n {\bf 1}$, 
and therefore
\be
 \frac{\partial X^{\prime }}{\partial X} \ = \ 
 h(t=1) \ = \  \sum_{n=0}^\infty  {1\over n!} h^{(n)} (t=0) 
 \ = \ \sum_{n=0}^\infty  {1\over n!}  (-1)^n  (\xi + a)^n {\bf 1}\ = \ 
 e^{-(\xi + a) } {\bf 1} \,.
 \ee
 In summary we have shown that 
 \be
\label{vmlvlvm99}
 \frac{\partial X^{\prime }}{\partial X} \ = \  
  e^{- (\xi + a)}  e^\xi \ = \  \Bigl( e^{- (\xi + a)}  \,{\bf 1} \Bigr) \,.
\ee
With this result we can readily write out the inverse matrix
 \be
\label{vmlvlvm99inv}
 \frac{\partial X}{\partial X'} \ = \  
   e^{-\xi}
    e^{\xi + a}  \ = \  \Bigl( {\bf 1} \, e^{- 
    \overleftarrow\xi  + a}   \Bigr) \,.
\ee
The first equality follows directly from (\ref{vmlvlvm99}), the second
by a calculation completely analogous to that above.  
Here, we have introduced the notation    
${\cal M} (  -\overleftarrow\xi  + a) \equiv
- (\xi {\cal M}) + {\cal M} a $ for the action of this operator on an arbitrary matrix ${\cal M}$. The expansion then gives
\be
 \frac{\partial X}{\partial X'}  \ = \   {\bf 1} + a +{1\over 2} a (  -\overleftarrow\xi  + a)
    + {1\over 3!}  a (  -\overleftarrow\xi  + a)^2 
    + {1\over 4!}  a (  -\overleftarrow\xi  + a)^3
    + {\cal O} (\xi^5)  \,.
\ee
Expanding the $\overleftarrow\xi$ action we find
  \be
\label{cbhuseful-alt-vmvs}
 \frac{\partial X}{\partial X'} \ = \
1+ a - {1\over 2} \xi a +{1\over 2} a^2 
+  {1\over 6} (\xi^2 a  - 2 (\xi a)a  - a\xi a + a^3)  + {\cal O}(\xi^4)\,. 
\ee

It is also of interest to find an expression for ${\cal F}$,
as defined in (\ref{inmatrixFvm}).  For this we need a formula
for $\bigl( {\partial X'\over \partial X}\bigr)^t$.  Using equation
(\ref{primepartialnormal-vcm}) one quickly notes that 
 \be
 \label{primepartialnormal-vcm-tr}
 \begin{split}
 \Bigl( \frac{\partial X^{\prime }}{\partial X}\Bigr)^t \ = \  & \ {\bf 1}-a^t
  +\frac{1}{2}a^t (\overleftarrow\xi+a^t)
  -\frac{1}{3!}a^t(\overleftarrow\xi+a^t)^2 
  +\frac{1}{4!} a^t(\overleftarrow\xi+a^t)^3 +\cdots  \\
  \ = \ & \ \Bigl( {\bf 1}\, e^{-(\overleftarrow\xi + a^t) } \Bigr) \ = \ 
  e^{-\xi} e^{\xi - a^t}  \,,
  \end{split} 
 \ee
where in the last step we used the second
equality in  (\ref{vmlvlvm99inv}) with $a\to -a^t$.
At this point it is useful to define a function ${\cal E}$ that appears 
both in (\ref{vmlvlvm99inv}) and (\ref{primepartialnormal-vcm-tr}).
We take
\be
\label{calEdefined}
\begin{split}
{\cal E} (k) \ \equiv\ &  \ e^{-\xi} e^{\xi + k} \ = \ 
\Bigl( {\bf 1} \, e^{- 
    \overleftarrow\xi  + k}   \Bigr)\,, \\
    = \ & 1+ k - {1\over 2} \xi k +{1\over 2} k^2 
+  {1\over 6} (\xi^2 k  - 2 (\xi k)k  - k\xi k + k^3)  + \ldots\;, 
    \end{split}
\ee
where we  made use of (\ref{vmlvlvm99inv}) and its expansion
(\ref{cbhuseful-alt-vmvs}).  We now have 
\be
\begin{split}
 \frac{\partial X}{\partial X'} \ =  \  {\cal E}(a) \,, 
 \qquad  
  \Bigl( \frac{\partial X^{\prime }}{\partial X}\Bigr)^t \ =  \ 
  {\cal E} (-a^t) \;. 
\end{split}
\ee
It follows that 
\be
\label{vmvmvm}
{\cal F} \ = \  {1\over 2} \Bigl( {\cal E}(a) {\cal E}(-a^t)+ 
{\cal E}(-a^t){\cal E}(a) \Bigr) \;. 
\ee
An expansion to cubic order in $\xi$ is now easily calculated.
We find
\be\label{Ftocubic}
\begin{split}
{\cal F} \ = \ & \,  1 + (a-a^t)  -  {1\over 2} \xi (a-a^t) + 
{1\over 2} (a-a^t)^2 \\
&+ {1\over 6} \xi^2 (a-a^t) 
- {1\over 3}\bigl(  \xi (a-a^t)\bigr) \,  (a-a^t) \\
&
- {1\over 6} (a-a^t) \xi (a-a^t)  + {1\over 6} (a-a^t)^3 
 \\
& -{1\over 12} \Bigl( (\xi a) a^t - a \xi a^t +a^2 a^t-a(a^t)^2\Bigr) + {\cal O}(\xi^4)\;. 
\end{split}
\ee
Comparing with (\ref{calEdefined})
we recognize that the first three lines fit precisely the cubic
expansion of ${\cal E} (a-a^t)$, and so we can write
\be
\label{Ftocubic-E}
{\cal F} \ = \  \,  {\cal E} (a-a^t)
 -{1\over 12} \Bigl( (\xi a) a^t - a \xi a^t +a^2 a^t-a(a^t)^2\Bigr) + {\cal O}(\xi^4)\,.
\ee

\subsection{Ordinary scalar and vector}\label{ordscaandvec}

Before turning to the generalized coordinate transformations let us 
review for scalars and vectors  the derivation of the finite gauge transformations as 
exponentials of ordinary Lie derivatives 
corresponding to ordinary diffeomorphisms associated with 
(\ref{newXprime}).

Consider the general situation of a field $\Psi$ whose
infinitesimal gauge transformation  is given by the action
of an operator ${\cal M}_\xi$ linear in the infinitesimal
gauge parameter $\xi$ but  field independent.  We write
\be
\Psi' (X) \ = \ \Psi(X) + {\cal M}_\xi \Psi (X) \,,
\ee   
or, more schematically,
\be
\Psi'  \ = \ \Psi + {\cal M}_\xi \Psi \,. 
\ee   
In order to construct a finite transformation with finite parameter
$\xi$ we define 
$\Psi (X; t)$ in such a way that $\Psi(X; t=0) = \Psi(X)$
and 
\be
\Psi (X; t + dt) \ = \  \Psi(X; t) + {\cal M}_{dt\hskip1pt \xi} \Psi(X; t)\,, 
\ee
which states that a change of parameter $dt$ is implemented
by a gauge transformation with parameter $dt\hskip1pt \xi$.
One can view $\Psi(X; t)$ as the gauge-transformed field obtained
for gauge parameter $t\xi$ and the fully transformed field is
$\Psi (X; t=1)$. Because of the linearity of ${\cal M}_\xi$
in $\xi$, the above equation implies that
\be
{d\Psi(X;t)\over dt~} \ = \ {\cal M}_\xi \Psi (X; t) \,.
\ee
Since ${\cal M}_\xi$ is field independent, we integrate this immediately
and find
\be
\Psi (X; t) \ = \  e^{t {\cal M}_\xi} \Psi (X; t=0)\,.
\ee
In conclusion, the fully transformed field $\Psi'(X) = \Psi (X; t=1)$ is 
given by
\be
\Psi'(X) \ = \  e^{{\cal M}_\xi} \, \Psi (X) \,. 
\ee
This is the desired large gauge transformation.

As a warmup let us consider the case of a scalar field.
Then the infinitesimal gauge transformation reads
\be
\phi'(X)  \ = \  \phi (X) + \xi^P\partial_P \phi(X) \ = \ \phi(X) + {\cal L}_\xi \phi (X) \,. 
\ee
Here ${\cal L}_\xi$ denotes the usual Lie derivative, and it is acting on the
scalar.  The above discussion implies that the large gauge transformation is given by
\be
\label{phi-large}
\phi'(X) \ = \  e^{{\cal L}_\xi}\, \phi(X)  \ = \  e^\xi\, \phi (X) \;, 
\ee  
since $\xi  =  \xi^M\partial_M$ coincides with the Lie derivative 
acting on a scalar. 
We now want
to show that this result follows from the basic transformation 
law
 \be\label{ScalarC}
 \phi'(X')  \ = \ \phi (X) \,, 
 \ee
 for the coordinate transformation (\ref{newXprime}).
 As written in (\ref{diffmc}) we have 
 \be
\label{diffmc-vm}
X' \ = \ e^{-\xi} X\  ~~\to ~~   e^\xi X' \ = \ X \,.
\ee
The last equation requires a little explanation.  The $\xi$
operator must act through the chain rule, as it involves $X$-derivatives.
The result is a function, as all derivatives must act on something.
We now use that for a general (analytic)  
 function $f$
 \be\label{generalfrel}
  (e^\xi f)(X^{\prime}) \ = \ e^\xi f(X')  e^{-\xi} \ = \ f(e^{\xi}X^{\prime}e^{-\xi}) \ = \ f(X)\,,
\ee
using  (\ref{Xprimenew}) and the logic that led to it.
Thus, $e^\xi$ acts on any (analytic) function by turning $X'$ into $X$. Therefore, 
 \be\label{functionprop}    
 e^\xi  \phi'(X') \ = \  \phi' (X)\,, 
 \ee
where here and henceforth we omit the parenthesis around $e^\xi  \phi'$.   
Therefore, using also the scalar property (\ref{ScalarC}) 
we have 
\be
 \phi'(X) \ = \  e^\xi  \phi' (X') \ = \ e^\xi  \,\phi (X) \ = 
 \ e^{{\cal L}_\xi}  \, \phi (X) \,, 
 \ee    
just as we had in (\ref{phi-large}).  This is what we wanted to show.

For a scalar density $\Phi (X)$ such as $e^{-2d}$ in (\ref{densitytrans}) we have
infinitesimally
\be
\label{infden}
\Phi'(X)  \ = \  \Phi (X) +  \partial_M \bigl( \xi^M \,\Phi )  \ = \  \Phi (X) + {\cal L}_\xi \Phi (X)\,,
\ee
where here ${\cal L}_\xi$ 
denotes a Lie derivative on the density.  This derivative, 
so defined to act on a density, satisfies the algebra
\be
\label{densityalgebra}
\big[ {\cal L}_{\xi_1}\,, {\cal L}_{\xi_2} \big] \ = \ - {\cal L}_{[\xi_1, \xi_2]} \,. 
\ee
This is the same algebra of diffeomorphisms that Lie derivatives satisfy acting
on arbitrary tensors.

\medskip
Let us now consider an ordinary vector field, whose
infinitesimal coordinate transformation takes the 
form\footnote{     
Ordinary and generalized vectors will be denoted by 
the same symbol $A_M$ and are recognized by the context.}
\be
A'_M   \ = \  A_M  +  \xi^K\partial_K A_M  + (\partial_M \xi^K) A_K\ = \
A_M + ({\cal L}_\xi A )_M\;. 
\ee
All fields here are evaluated at the common argument $X$.
In index free notation we have
\be
A'   \ = \  A  +  \xi  A +  a\, A\ = \
A + {\cal L}_\xi A \,,
\ee
which shows that, on the vector, we can view ${\cal L}$ as the matrix operator
\be
{\cal L}_\xi \ = \  \xi + a \,.
\ee
It then follows that the large coordinate transformation
of the vector is given by
\be
\label{largectvec}
A'  \ = \  e^{{\cal L}_\xi}  A  \ = \  e^{\xi + a}  A \,. 
\ee

We now compare with the large gauge transformation derived
from the usual coordinate transformation of a vector, 
\be
A'_M(X') \ = \  {\partial X^N \over \partial X'^M}  A_N (X) 
\quad \to \quad  
A'(X') \ = \  {\partial X \over \partial X'}  A (X) \,.
\ee
Following 
(\ref{functionprop})  we write
\be
A' (X) \ = \ e^{\xi} A'(X')  = e^{\xi} \,{\partial X \over \partial X'}  A(X)\,,
\ee
or, leaving out the common argument, 
\be
A'  \ = \  e^{\xi}\,  {\partial X \over \partial X'}   \, A\,.
\ee
The above partial derivatives were calculated in 
(\ref{vmlvlvm99inv}). Using them we have
\be
A'  \ = \  e^{\xi} \bigl( e^{-\xi}
    e^{\xi + a} \bigr)  \, A \ = \  e^{\xi + a}  \, A \,,
\ee
in agreement with (\ref{largectvec}).

\subsection{Generalized vector and reparameterized
diffeomorphisms}

The case of a generalized scalar is no different from
the ordinary scalar.  For generalized vectors, however,
the situation is quite different.
The infinitesimal transformation of a generalized
vector is given by the generalized Lie derivative, 
\be
A'_M   \ = \  A_M  +  \xi^K\partial_K A_M  + \bigl(
\partial_M \xi^K  - \partial^K \xi_M) A_K\ = \
A_M + (\widehat {\cal L}_\xi A )_M\;. 
\ee
In index-free notation we have
\be
A'   \ = \  A +  \xi A + \bigl( a - a^t\bigr) A\ = \
A + \widehat {\cal L}_\xi A \,,
\ee
which shows that on a generalized vector we can view
the generalized Lie derivative as the operator
\be
\widehat{\cal L}_\xi  \ = \  \xi + a - a^t \,. 
\ee
It follows that the large  
gauge transformation of the
vector is then given by 
\be
\label{largenveclie}
A' (X) \ = \  e^{\widehat {\cal L}_\xi} \, A\ = \  e^{\xi + (a-a^t)} \, A\,. 
\ee

We now must compare with the transformation (\ref{finiteF}) we 
postulated.  Following the steps that are by now familiar, we have
\be
A'_M (X) \ = \ e^{\xi} A'_M(X')  = e^{\xi} {\cal F}_M{}^N A_N(X)\;, 
\ee
or in matrix notation
\be
A' (X) \ = \  e^{\xi}\, {\cal F} \, A \,. 
\ee
Equality  with (\ref{largenveclie}) would require
\be
\label{tentative}
{\cal F} \ = \ e^{-\xi} e^{\xi + (a-a^t)} \;\; ? 
\ee
Using the definition (\ref{calEdefined}) we are thus asking if 
\be
\label{tentative-vm}
{\cal F}  \ = \ {\cal E} (a-a^t)  \;\; ?  
\ee
In here, ${\cal F}$ is calculated using the   
diffeomorphism $X' = \exp(-\xi) X$ and its definition
(\ref{inmatrixFvm}).  The result to cubic order was
given in (\ref{Ftocubic}) and (\ref{Ftocubic-E}).  
We found there that the above relation
holds up to quadratic order, but not to cubic order: 
\be\label{cubicrticdiff-vm}
   {\cal F}
 \ =  \ {\cal E}(a-a^t)  - \Delta{\cal F} \,,
 \ee
 where the correction $\Delta {\cal F}$ is given by
\be\label{quarticdiff}
  \Delta{\cal F} 
 \ =  \  
   {1\over 12} 
   \Bigl( (\xi a) a^t - a \,\xi a^t +a^2 a^t-a(a^t)^2\Bigr) + 
   {\cal O} (\xi^4)\,.
 \ee

This is an apparent failure of consistency.  But there is some
freedom in double field theory that is not available in ordinary
field theory.  We can use that freedom to alter the parameterization
of the diffeomorphism in such a way that the vector field transformations
work out.  In doing so we must be careful not to spoil the already
achieved agreement for the scalar field.

The diffeomorphism we have been considering so far is
 \be
  X^{\prime M}_\xi \ \equiv \ e^{-\xi^{P}\partial_{P}}X^{M} \,,
 \ee
where we have added the subscript $\xi$ to emphasize the
role of this parameter. Now we consider a different 
diffeomorphism 
 \be
 \label{xtheta0}
  X^{\prime M}_\Theta \ \equiv \ e^{-\Theta^{P}\partial_{P}}X^{M} \ = \ 
  X^{M}-\Theta^{M}+ {1\over 2}  \Theta^P \partial_{P}\Theta^M -
   {1\over 3}  \Theta^P \partial_{P}\, \Theta^K \partial_{K}\Theta^M
 +  {\cal O}(\Theta^4) \;.  
 \ee
 We are to design the new diffeomorphism -- or equivalently to fix
 $\Theta(\xi)$ -- so  
  that ${\cal F}_\Theta$,
 given by
\be
\label{calFTheta}
 {\cal F}_{\Theta} \ \equiv  \  \    {1\over 2} \,
  \Bigl( \,  \frac{\partial X}{\partial X'_\Theta}\,
  \Bigl(\frac{\partial X^{\prime}_\Theta}{\partial X}\Bigr)^t
  +    \Bigl(\frac{\partial X'_\Theta}{\partial X}\Bigr)^t\,
  \frac{\partial X}{\partial X'_\Theta}\, \Bigr)  \;, 
  \ee
satisfies the requisite relation
\be
\label{FTheta=E}
{\cal F}_\Theta = {\cal E}(a-a^t)\,,
\ee
that guarantees that ${\cal F}_\Theta$ generates the same
transformation as the exponential of the generalized
Lie derivative.  
In here we will achieve the above equality up to terms
of order $\xi^3$ and in the appendix   
we extend the result to order $\xi^4$.

We now consider the case when $\Theta$ equals $\xi$ to leading
order but has higher order corrections.  Since $\Delta {\cal F}$ is
cubic in $\xi$  we have no use for quadratic
corrections and we write 
\be\label{Thetainzivm}
\Theta^M \ = \ \xi^M   - \delta_3^M (\xi)  + \ldots
\ee
The subscript in $\delta$  
indicates that this term is cubic in $\xi$.
 We will also assume that in $\delta_3^M$ the
index $M$ is carried by a derivative. Schematically,
\be
\Theta^M \ = \ \xi^M +  \sum_{i}    
\rho_{i}\,
\, \partial^M \chi_{i}\,,
\ee
with $\rho_i$ and $\chi_i$ functions of $\xi$. 
Because of the strong constraint, 
 the action of $\Theta^P \partial_P$  on fields (like $\xi$, or $\Theta$, but
not $X$), reduces 
to the action of 
$\xi^P\partial_P$:\footnote{We note that this modification is consistent 
with the transformation of a density like the dilaton, which is unmodified compared to 
ordinary geometry, see (\ref{densitytrans}), because by the 
strong constraint the extra 
term $\partial_M\xi^M$ in the transformation rule is also unchanged when replacing 
$\xi$ by $\Theta$.}
\be
\Theta^P\partial_P  (\,\hbox{fields}\,)  \ = \  \xi^P \partial_P (\,\hbox{fields}\,) \;. 
\ee
Applied to (\ref{xtheta0}) this gives
\be
 \label{xtheta-vmvbvb}
 \begin{split}
  X^{\prime M}_\Theta \ \equiv & \ \ e^{-\Theta^{P}\partial_{P}}X^{M} \ = \ 
  X^{M}-\Theta^{M}+ {1\over 2}  \xi^P \partial_{P}\Theta^M
  - {1\over 3!}  \xi^P \partial_{P}\xi^K \partial_K\Theta^M
  +  
  {\cal O}(\xi^4) \\
 \ =& \   \ X^{M}-\Theta^{M}+ {1\over 2}  \xi \Theta^M 
- {1\over 3!} \xi \xi \Theta^M   + 
  {\cal O}(\xi^4)\;.    
\end{split}
 \ee
On a scalar the new diffeomorphism results in the same large
coordinate transformation.  Since
\be
X'_\Theta \ = \   e^{-{\Theta}^P\partial_P }X  \quad \to \quad  X \ = \ 
 e^{{\Theta}^P\partial_P } X'\,,   
 \ee
we have, as before (see the discussion starting with
(\ref{diffmc-vm}) and leading to (\ref{functionprop})), 
\be
  \phi^{\prime}(X) \ = 
   e^{{\Theta}^P\partial_P }\phi'(X') \ = \  e^{{\Theta}^P\partial_P }\phi(X)
  \ = \ e^{\xi^{P}\partial_{P}}\phi(X)
 \ = \ e^{\widehat{\cal L}_\xi}\phi(X)\;.
 \ee

We now aim to compute ${\cal F}_\Theta$. 
First, using (\ref{Thetainzivm}) and (\ref{xtheta-vmvbvb}), we can  write the relation between the two $X$'s as 
\be
X'^M_{\Theta}  \ = \  X'^M_{\xi}  + \delta_3^M + {\cal O} (\xi^4)\;.
\ee
We then have 
\be
\begin{split}
\label{ghghx}  
{\partial X'_\Theta\over \partial X} \ = \ & \ 
{\partial X'_\xi\over \partial X}  + \Delta_3\,, ~~~~
({\Delta}_3)_Q{}^M  \ \equiv \ \partial_Q  \delta_3^{~M} \ = \ (\partial \delta_3)_Q{}^M \,, \\
{\partial X\over \partial X'_\Theta} \ = \ &  \ 
{\partial X\over \partial X'_\xi}  - \Delta_3 \,.
\end{split}
\ee
We use the definition (\ref{calFTheta}) to find that   
\be
\label{fthetafxi}   
{\cal F}_{\Theta} 
 \ = \  {\cal F}_{\xi}   +\Delta_3^t -  \Delta_3  \,.
\ee
In this light we have from (\ref{cubicrticdiff-vm}) 
\be
{\cal F}_{\Theta} \ =   \ {\cal E}(a-a^t)  - \Delta{\cal F}  +\Delta_3^t -  \Delta_3\,.
\ee
We are to design the new diffeomorphism so that ${\cal F}_\Theta$
is equal to ${\cal E}(a-a^t)$, and therefore we must
find a  $\Theta (\xi)$ 
for which
\be
\label{mustshow-vm}
\Delta_3^t -  \Delta_3 \ = \   \Delta{\cal F} \,. 
\ee
We claim that 
 $\Theta$ is given by  
\be
\label{Thetatoquartic}
\Theta^M \ = \ \xi^M  + {1\over 12}  \, (\xi \xi^L) \partial^M \xi_L 
+ {\cal O}(\xi^4) \,, \ee    
or equivalently
\be
  \delta_3 \ = \ -  {1\over 12}  \, (\xi \xi^L) \partial^M \xi_L   \,.
\ee
We confirm this quickly. 
The definition of $\Delta_3$ in (\ref{ghghx}) gives
\be\label{ourcase}
\Delta_3 \ = \ \partial \delta_3 \ = \ -{1\over 12}  \bigl( (\xi a)a^t  + a^2 a^t \bigr)  
-{1\over 12}  (\xi \xi^L)  \partial \partial  \xi_L \,,  \ee
where the matrix indices are carried by the partial derivatives $\partial \partial$ in the second term.  Moreover
\be
\Delta_3^t -  \Delta_3  \ = \   
{1\over 12} \Bigl( (\xi a) a^t  + a^2 a^t - a\xi a^t
- a (a^t)^2 \Bigr)\,.
 \ee
This coincides exactly with $\Delta {\cal F}$ in (\ref{quarticdiff}).
Thus equation (\ref{mustshow-vm}) holds and we have completed 
the verification of (\ref{FTheta=E}) to order $\xi^3$:
\be
\label{FTheta=Exx}
{\cal F}_\Theta = {\cal E}(a-a^t)+ {\cal O}(\xi^4)\,. 
\ee
In Appendix A  we carry the computation to quartic
order and show that the above $\delta_3$ suffices to generate
the terms that must be cancelled.  Thus the above 
actually holds with ${\cal O} (\xi^5)$. We expect that there will
be a need for a correction $\delta_5$ to quintic order.

\section{Composition of generalized coordinate transformations}

In this section we study the composition of gauge transformations.
As we will argue, our previous results that relate large gauge
transformations to exponentials of Lie derivatives guarantee the
existence of a composition law.  This is true both for the ordinary
and for the generalized case.  It will also become clear here that
in the generalized case the composition of the underlying 
coordinate transformations is exotic.

\subsection{Facts on composition}
To begin we consider 
two diffeomorphisms  
\be
\begin{split}
X'  \ = \ &  \  e^{-{\xi_1}(X)} X\,,\\
X''  \ = \ &\  e^{-{\xi_2}(X')} X'\,.
\end{split}
\ee
We also consider a 
diffeomorphism from $X$ to $X''$
\be
X''  \ = \  e^{- \xi_{12}(X)} 
X\,.
\ee
If this diffeomorphism is induced by the composition of the previous
two diffeomorphisms  
we have 
$X'' = e^{- \xi_{2}(X')} e^{-{\xi_1}(X)} X$ and therefore
\be
\label{vm}
 e^{-\xi_{12}(X)}  \ = \ e^{-{\xi_2}(X')}\, e^{-{\xi_1}(X)}\,.
\ee
In order for the argument of $\xi_2$ to become $X$ we multiply
the right hand side by unity, expressed as $e^{-\xi_1(X)} e^{\xi_1(X)}$: 
\be
\label{sgbvg}
 e^{- \xi_{12}(X)} \ =  \ e^{-\xi_1(X)} ~\bigl( e^{\xi_1(X)}
 e^{-{\xi_2}(X')}\, e^{-{\xi_1}(X)}\bigr) \,.
 \ee
Recall from (\ref{generalfrel}) that  
$e^{\xi_1} f(X') e^{-\xi_1} = f(X)$, for any regular function 
$f(X)$.  Using this we find the relations
\be
\begin{split}
\label{sgbvgx}
 e^{-\xi_{12}(X)} \ = \ &  \ e^{-{\xi_1}(X)}\,e^{-{\xi_2}(X)} \,, \\
  e^{ \,\xi_{12}(X)} \ =  \ & \,e^{\,{\xi_2}(X)}\, e^{\,{\xi_1}(X)}\,, 
 \end{split}
 \ee
where the second line is obtained by taking 
the inverse of the first
line.  We are now in a position to use the 
Baker-Campbell-Hausdorff (BCH) 
relation to write an explicit expression for $\xi_{12}$:
 \be
 \label{sgxi12vg}
 \xi_{12} \ = \ \xi_2 + \xi_1 + {1\over 2} \big[\xi_2, \xi_1 \big] 
 + {1\over 12} \bigl(  \big[\xi_2,  \big[\xi_2, \xi_1 \big] \big] +  \big[\xi_1,  \big[\xi_1, \xi_2 \big] \big] \bigr) + \ldots
 \ee
For our applications here we will just need this formula to quadratic
order in $\xi$:
\be
\label{xi12lead9}
\xi_{12} \ = \   \xi_1 + \xi_2  -  {1\over 2} \bigl[\,\xi_1 \,,\, \xi_2
\, \bigr] + {\cal O}(\xi^3)\,. 
\ee

A useful alternative picture of the situation involves the 
Lie derivative operator ${\cal L}_\xi$.  The key properties of this
operator are its linearity in $\xi$ and the commutator
$[{\cal L}_{\xi_1} , {\cal L}_{\xi_2} ]  =  - {\cal L}_{[\xi_1, \xi_2]}$.\footnote{Let us note 
that here and 
below  
(see eq.~(\ref{ACTing})), we employ the convention 
that the (generalized) Lie derivatives act on the fields first. The opposite convention 
according to which the Lie derivatives are operators acting equitably on everything on the right 
leads to a different sign in the commutator of Lie derivatives.} 
We can combine
the exponentials of two such operators as follows
\be
\label{vmsg}
e^{{\cal L}_{\xi_1(X)} } e^{{\cal L}_{\xi_2(X)} } \ = 
\ e^{{\cal L}_{\,{\xi}_{12}(X)} } \,,
\ee
where we claim that  $\xi_{12}$ is the one determined above.
To see this, let the above operator equation act on a scalar field $S$.
On a scalar the Lie derivative acts just like the vector
operator: ${\cal L}_\xi S=  \xi S$,  and we therefore  get
\be\label{ACTing}
e^{{\cal L}_{\xi_1(X)} } e^{\xi_2(X) }  S \ = \
 e^{\xi_2(X) }  e^{{\cal L}_{\xi_1(X)} }S \ = \
 e^{\xi_2(X) }  e^{\xi_1(X)}S \ = 
\ e^{ \xi_{12}(X)} S\,,
\ee
consistent with (\ref{sgbvgx}).
We can also explicitly combine the operators on the left-hand
side of (\ref{vmsg}) using BCH:  
\be  
e^{{\cal L}_{\xi_1(X)} } e^{{\cal L}_{\xi_2(X)} }\, =\, 
e^{ {\cal L}_{\xi_1} + {\cal L}_{\xi_2} + {1\over 2} [ {\cal L}_{\xi_1}\,, 
{\cal L}_{\xi_2} ] + \ldots} 
 =   \  e^{ {\cal L}_{\xi_1} + {\cal L}_{\xi_2}  -{1\over 2} {\cal L}_{[\xi_1, \xi_2]} + \ldots} 
 = \  e^{ 
 {\cal L}_{\xi_1+\xi_2  -{1\over 2} [\xi_1, \xi_2]  + \ldots}    } \,.
\ee
This, of course, gives the same determination of $\xi_{12}$.

\medskip
In the generalized case the coordinate transformations
become subtle to handle, but the analogy to Lie derivatives
holds.  Thus, in view of (\ref{vmsg}) we now consider the corresponding
exponentials of generalized Lie derivatives, 
\be
\label{wkdflkd}
e^{\widehat{\cal L}_{\xi_1(X)} } e^{\widehat{\cal L}_{\xi_2(X)} } \ = \ e^{\widehat{\cal L}_{\, {\xi}^c_{12}(X)} } \,.
\ee
As for the BCH  
relation, the
only difference with ordinary Lie derivatives is that  the
commutator of generalized Lie
derivatives gives a generalized Lie derivative with parameter
equal to (minus) the {\em Courant} bracket of the parameters.
It follows that the parameter $\xi_{12}^c$ written above is in fact
given by the same formula (\ref{sgxi12vg})  that gives $\xi_{12}$ but this time
using the Courant bracket:
 \be
 \label{sgxi12vgc}
 \xi_{12}^c \ = \ \xi_2 + \xi_1 + {1\over 2}  \big[\xi_2, \xi_1 \big]_c 
 + {1\over 12} \bigl(  \big[\xi_2,  \big[\xi_2, \xi_1 \big]_c \big]_c +  \big[\xi_1,  \big[\xi_1, \xi_2 \big]_c \big]_c \bigr) + \ldots\,.
 \ee
It is important to clarify the notation in the above formula.  In the generalized theory, due to the strong constraint,  it is not 
synonymous to speak of the
components $A^M$ of a vector, or the vector operator $A^M\partial_M$.
The above equation  must  be thought of as an 
equation for {\em components}:  
\be
\label{sgomg}
 (\xi_{12}^c)^M \ = \ \xi_2^M + \xi_1^M + {1\over 2}  \big[\xi_2, \xi_1 \big]_c^M 
 +  \ldots \,.
\ee
This distinction is relevant: while the vector 
components $(\xi_{12}^c)^K$ and $(\xi_{12})^K$ are 
not equal, we claim that the strong constraint implies
 the equality of the corresponding 
 vector operators
\be
\label{kugg}
 (\xi_{12}^c)^K \partial_K
\ = \ (\xi_{12})^K \partial_K \,.
\ee
We can write this as  $\xi_{12}^c  =  \xi_{12}$ when no
confusion is possible, but recalling that the vectors do not
have the same components. 
In order to prove (\ref{kugg}) we first 
recall that  the C-bracket (\ref{Cbracket})  
differs from the Lie bracket by 
a  vector whose  index is carried by a derivative:
\be
\label{kuvb}
[A, B]_c^K \ = \  [A, B]^K  +  (\cdots \partial^K \cdots)\,,
\ee
where the expressions indicated by dots are not presently relevant.
It then follows by the strong constraint that 
\be
\label{kuvg}
[ A, B]_c^M \partial_M  \ = \ [A, B]^M \partial_M\,.
\ee
Next we verify the following relation for Lie brackets:
\be
\label{kubvg}
\bigl[ ~ \chi\,,  (\cdots \vec{\partial} \cdots ) \bigr] \ = \  (\cdots  \vec{\partial} \cdots)\,.
\ee
This states that the Lie bracket of an arbitrary vector with
a vector whose index is carried by a derivative is again
a vector whose index is carried by a derivative.  This is
readily verified computing the $M$-th component of the
following commutator:
\be
\bigl[ ~ \chi\,,  \rho \,\vec{\partial}\,\eta \,  \bigr]^M \ = \ \chi^K \partial_K(\rho\, \partial^M \eta) - (\rho\partial^K \eta) \partial_K \chi^M \ = \  \chi^K \partial_K(\rho\, \partial^M \eta)\,,
\ee
and, as claimed, the right hand side is a vector whose index is carried by a
derivative.  We can now see that a double nested C-commutator also
reduces to a Lie commutator:
\be
\begin{split}
\bigl[ \,\chi_1\,, [\chi_2, \chi_3]_c \,]_c^M \partial_M \ = \ & \  
\bigl[ \,\chi_1\,, [\chi_2, \chi_3]_c \,]^M \partial_M\\
=\ & \  
\bigl[ \,\chi_1\,, [\chi_2, \chi_3] \,]^M \partial_M
+ \bigl[ \,\chi_1\,,  (\cdots \vec{\partial} \cdots) \,]^M \partial_M\,,
\end{split}
\ee
using (\ref{kuvg}) and then (\ref{kuvb}).  We now use (\ref{kubvg}) to conclude that, as claimed
\be
\bigl[ \,\chi_1\,, [\chi_2, \chi_3]_c \,]_c^M \partial_M \ =
\bigl[ \,\chi_1\,, [\chi_2, \chi_3] \,]^M \partial_M
+(\cdots {\partial}^M \cdots)  \partial_M \ = \ 
\bigl[ \,\chi_1\,, [\chi_2, \chi_3] \,]^M \partial_M\,.
\ee
It is now easy to make an inductive argument to show
that 
\be
\bigl[ \,\chi_1\,, [\chi_2, [\chi_3\, \cdots  [\chi_{n-1}\,, \chi_n]_c ]_c
\ldots ]_c \,]_c^M \partial_M\ = \ \bigl[ \,\chi_1\,, [\chi_2, [\chi_3\, \cdots  [\chi_{n-1}\,, \chi_n] \,]
\ldots ] \,]^M \partial_M\,.
\ee
Indeed, in such an argument
one may assume that all commutators are Lie except
for the most nested one. Then one uses (\ref{kuvb}) 
for this commutator to get the
desired term with all brackets of Lie type and an extra term
where that most nested commutator is replaced by
a vector with index carried by a derivative.  Then successive application of (\ref{kubvg}) gives the desired result.  Having
shown this, and given the form of $\xi_{12}^c$ in (\ref{sgxi12vgc}), we
see that (\ref{kugg}) is true.

\subsection{General argument for composition}\label{genargcomp}

For the ordinary vector we wrote
\be
\label{traord}
A'(X') \ = \  {\cal G}(X', X) A (X)\,,
\ee
and with the diffeomorphism 
\be
X' \ = \ e^{-\xi} X\,,  
\ee
we found that  
\be
\label{trivi}
G(X', X) \ = \ {\partial X\over \partial X'}\,  ,   
\ee
can be written as 
\be
\label{gxx'formula}
{\cal G} (X', X) \ = \ e^{-\xi} \, e^{\,\xi + a} \;.   
\ee
Moreover, with these results we also found that 
transformation (\ref{traord}) implies \be
A'(X) \ = \ e^{{\cal L}_\xi}  A(X)  \ = \ e^{\,\xi + a}  A\,.
\ee
As indicated above, acting on vectors, 
${\cal L}_\xi  = \xi  + a$.
Therefore,  when the operators in (\ref{vmsg}) are acting on a vector we have
\be
\label{sgvor}
e^{\xi_2 + a_2} \,  e^{\xi_1 + a_1} \ = \ e^{\xi_{12} + a_{12}} \,.
\ee
We now verify 
that the composition property 
of ${\cal G}$, 
\be
\label{comp-=ord}
{\cal G} (X'' , X') \, {\cal G} (X' , X) \ = \  {\cal G} (X'', X)\,, 
\ee
is a consequence of (\ref{sgvor}).\footnote{Eqn.~(\ref{comp-=ord})  follows directly from (\ref{trivi}), but such derivation is not available for the generalized case.}  Given (\ref{gxx'formula}) the above equation
requires that   
\be
 e^{-{\xi}_2(X') } \, e^{\,\xi_2(X')  + a'_2}  ~  e^{-{\xi}_1(X) } \, e^{\,\xi_1(X)  + a_1}  \ = \ 
  e^{-{\xi}_{12}(X) } \, e^{\,\xi_{12}(X)  + a_{12}} \,.
\ee
 Let us show  
 that this gives us 
 (\ref{sgvor}).
 Acting with $e^{\,{\xi}_{12}(X) }$
\be
e^{{\xi}_{12}(X) }  e^{-{\xi}_2(X') } \, e^{\,\xi_2(X')  + a'_2}   ~ e^{-{\xi}_1(X) } \, e^{\,\xi_1(X)  + a_1}  \ = \ 
  \, e^{\,\xi_{12}(X)  + a_{12}} \,.
\ee
Then using (\ref{vm})
\be
\label{vmvb}
 e^{{\xi_1}(X)} \, e^{\,\xi_2(X')  + a'_2}\,    e^{-{\xi}_1(X) }\,  \, e^{\,\xi_1(X)  + a_1}  \ = \ 
  \, e^{\,\xi_{12}(X)  + a_{12}} \,.
\ee
The first three factors on the left-hand side give\footnote{Note that $a'_2$ goes to
$a_2$ because
$(a'_2)_Q{}^P  = \partial'_Q \xi_2(X')^P  =  h(X')_Q{}^P$ is ultimately
a function of $X'$ so that
$
e^{\xi_1} (a'_2)_Q{}^P e^{-\xi_1}  =  
e^{\xi_1}  h(X')_Q{}^P e^{-\xi_1}  =    h(X)_Q{}^P
=   \partial_Q \xi_2(X)^P    =  (a_2)_Q{}^P$.} 
the factor $ e^{\xi_2(X) + a_2}$.
Thus (\ref{vmvb}) becomes
\be
\e^{\xi_2(X) + a_2}
 ~ \, e^{\,\xi_1(X)  + a_1}  \ = \ 
  \, e^{\,\xi_{12}(X)  + a_{12}} \,,
\ee
which is identical to (\ref{sgvor}). 

\medskip
We can now turn to the generalized case.
The composition law on a scalar is no different from that in ordinary
geometry and holds as in that case.  For the scalar density 
we have that the Lie derivatives ${\cal L}_\xi$ considered in 
(\ref{infden}) lead to 
\be
\label{wkdflkd9999}
e^{{\cal L}_{\xi_1(X)} } e^{{\cal L}_{\xi_2(X)} } \ = \ e^{{\cal L}_{\, {\xi}_{12}(X)} } \,.
\ee
But for the scalar density (or the scalar) Lie derivatives take the same form as generalized
Lie derivatives, so we have
\be
\label{wkdflkd99998}
e^{\widehat{\cal L}_{\xi_1(X)} } 
e^{\widehat{\cal L}_{\xi_2(X)} } \ = \ e^{\widehat{\cal L}_{\, {\xi}_{12}(X)} } \,.
\ee
Moreover, acting on a scalar density (or a scalar) any contribution to $\xi^M$ 
of the form $\cdots \partial^M  \cdots$ will vanish on $\widehat{\cal L}_\xi$.  Thus
by virtue of (\ref{kugg}) we can replace $\xi_{12}$ by $\xi_{12}^c$ in the above, finding 
that on a scalar density we have 
\be
\label{wkdflkd99xcv}
e^{\widehat{\cal L}_{\xi_1(X)} } e^{\widehat{\cal L}_{\xi_2(X)} } \ = \ e^{\widehat{\cal L}_{\, {\xi}^c_{12}(X)} } \,.
\ee
The integration of the infinitesimal transformation of the scalar density leads to
(\ref{densitytrans}) 
and by the above argument, such transformation must be consistent
with composition, as expressed in the generalized case by the equation above.
We verify this explicitly at the end of section~\ref{testingcomposition}.

Let us now consider the 
large transformation of the vector field is represented by the 
relation
\be
\label{gkiu}
A'(X') \ = \  {\cal F}(X', X) A (X)\,,
\ee
with 
\be
\label{triv}
{\cal F}(X', X) \ = \ {1 \over 2} \Bigl( {\partial X\over \partial X'}
{\partial X'\over \partial X}^t + {\partial X'\over \partial X}^t{\partial X\over \partial X'}\Bigr) \,.
\ee
We have already 
shown  that 
\be
X' \ = \ e^{-\Theta(\xi)} X \,,   ~~~\Theta (\xi) =  \xi + {\cal O}(\xi^3) \,,
\ee
leads to 
\be
{\cal F} (X', X) \ = \ e^{-\xi} \, e^{\,\xi + k} \,,    
~~~~ k \ = \ a - a^t\, ,  
\ee
at least to ${\cal O}(\xi^5)$.  
Moreover, this and (\ref{gkiu}) imply that 
\be
A'(X) \ = \ e^{\widehat{\cal L}_\xi}  A(X)  \ = \ e^{\,\xi + k}  A\,.
\ee
Note that on generalized vectors 
$\widehat{\cal L}_\xi =  \xi +k$.     
 It now follows that the composition (\ref{wkdflkd}) of exponentials
 of generalized Lie derivatives, applied to generalized
vectors,  gives
\be
\label{vmvm}
 e^{\xi_2 (X) + k_2}  e^{\xi_1(X) + k_1 }  \ = \  e^{{\xi}^c_{12} (X) + k_{12}^c}\,.
\ee
We claim that composition of 
${\cal F}$ holds in the following sense:
\be
\label{vgsg}
 e^{-{\xi}_2(X') } \, e^{\,\xi_2(X')  + k'_2}  ~  e^{-{\xi}_1(X) } \, e^{\,\xi_1(X)  + k_1}  \ = \ 
  e^{-{\xi}_{12}^c(X) } \, e^{\,\xi_{12}^c(X)  + k^c_{12}} \,.
\ee
This means that 
\be
\label{comp-=ordf}
{\cal F} (X'' , X') \, {\cal F} (X' , X) = {\cal F} (X'', X)\,, 
\ee
where  the ${\cal F}$ on the right-hand side  is  
 built
from $X'' = e^{-\Theta (\xi_{12}^c)} X$.
To prove (\ref{vgsg})  we first multiply it 
by  $e^{\xi_1(X)} e^{\xi_2(X')}$ to get
\be
\label{lcksgvg}
 e^{\xi_1(X)}  \, e^{\,\xi_2(X')  + k'_2}  ~  e^{-{\xi}_1(X) } \, e^{\,\xi_1(X)  + k_1}  \ = \ 
e^{\xi_1(X)}     e^{\xi_2(X')}    e^{-\xi_{12}^c(X) } 
\, e^{\,\xi_{12}^c(X)  + k^c_{12}} \,~.
\ee
Consider the first three factors on the above right-hand side.
Given (\ref{kugg}) we can replace $\xi_{12}^c$ by $\xi_{12}$ 
(since here they are  
operators) and then use (\ref{vm}) 
to find that these factors
 give the unit matrix:
\be
e^{\xi_1(X)}     e^{\xi_2(X')}    e^{-\xi_{12}^c(X) } \ = \ 
e^{\xi_1(X)}     e^{\xi_2(X')}    e^{-\xi_{12}(X) } \ = \ {\bf 1} \,.
\ee
On the left-hand side of (\ref{lcksgvg}) we see that the first and third factor implement
the change $X'\to X$ on the second factor.  All in all (\ref{lcksgvg})
becomes
\be
\label{lcksgvg99}
  e^{\,\xi_2(X)  + k_2}  \, e^{\,\xi_1(X)  + k_1}  \ = \ 
\, e^{\,\xi_{12}^c(X)  + k^c_{12}}\,. 
\ee
This is indeed identical to (\ref{vmvm}), as we wanted to show.
Note that the above right-hand side is also equal to 
$e^{\,\xi_{12}(X)  + k^c_{12}}$, but  $k^c_{12}$ is built from the
{\em components}  $(\xi_{12}^c)^M$, and therefore cannot 
be traded for $k_{12}$ which is build from the components 
$(\xi_{12})^M$.

\subsection{Testing composition} \label{testingcomposition}

In this section we test explicitly the composition rules.
This provides a confirmation of the arguments presented
above and is simply a welcome check on the formalism.
While the confirmation to be done is certainly not novel
in the ordinary geometry case,  the notation to
be introduced will help the treatment of the generalized case.

For the three parameters $\xi_1, \xi_2,$ and $\xi_{12}$ we introduce
the matrices $a_1, a_2,$ and $a_{12}$ as 
the analogs of the 
matrix $a(\xi)_P{}^Q = \partial_P \xi^Q$:
 \be\label{ABCdef}
  (a_1)_{M}{}^{N} \ \equiv \ \partial_{M}\xi_1^{N}\;, \quad (a_2)_{M}{}^{N} \ \equiv \ \partial_{M}^{\prime}\xi_{2}^{N}(X^{\prime})\;, \quad
  (a_{12})_{M}{}^{N} \ \equiv \ \partial_{M}\xi_{12}^{N}(X)\;.
\ee
The composition law (\ref{comp-=ord}) then requires
\be
\label{compowish}
  {\cal G}(\xi_2)\,{\cal G}(\xi_1) \ = \ {\cal G}(\xi_{12})\,,
 \ee
where we have denoted the ${\cal G}$ in terms of the parameter
that generates the corresponding transformation.  This equation
must determine $\xi_{12}$, and we expect that this is the
$\xi_{12}$ obtained before.
 
Recalling that ${\cal G} = \frac{\partial X}{\partial X'}$
and making use of 
(\ref{cbhuseful-alt-vmvs}) we can write, to quadratic order, 
 \be\label{oldF}
  {\cal G}(\xi_1) \ = \  {\bf 1}+a_1 -\frac{1}{2}{\xi_1}a_1
  +\frac{1}{2}a_1 a_1+\cdots \;.    
   \ee
Using (\ref{oldF}) for the other two $\xi$'s, we quickly
find that (\ref{compowish}) requires, to quadratic order, 
 \be
  {\bf 1}+a_1+a_2 -\frac{1}{2}{\xi_1}a_1
    -\frac{1}{2}{\xi_2}a_2
    +a_2a_1 +\frac{1}{2}\big(a_1a_1+a_2 a_2\big) 
    \ = \ {\bf 1}+a_{12}-\frac{1}{2}{\xi_{12}}a_{12}
    +\frac{1}{2}a_{12}a_{12}\;. 
 \ee
To linear order this requires $a_{12} = a_1+ a_2$. 
Writing $a_{12} = a_1 + a_2 + \delta$ one readily 
determines $\delta$ and concludes  that
the above equation is satisfied if
 \be\label{C}
  a_{12} \ = \ a_1+a_2
  -\frac{1}{2}[a_1,a_2]+\frac{1}{2}({\xi_1}a_2+{\xi_2}a_1)\;.
 \ee
We now calculate $a_{12}$, with 
$\xi_{12}$ given in (\ref{xi12lead9})  and will show
that indeed the above $a_{12}$ arises.  We begin with
 \be\label{CStep}
  (a_{12})_{M}{}^{N} \ = \ \partial_{M}\xi_{12}^{N} \ = \ \partial_M\big(\xi_1^{N}+\xi_2^{N}
  -\frac{1}{2}\xi_1^{P}\partial_{P}\xi_2^{N}+\frac{1}{2}\xi_2^{P}\partial_P \xi_1^{N}\big) \;. 
   \ee
In here we will have to evaluate
the derivative $\partial_{M}\xi_2^{N}(X)$
which is closely related to $a_2$.  The relation to quadratic order
is readily found,  
\be
\partial \xi_2 (X)  \ = \ 
e^{\xi_1} \, \partial' \xi_2 (X') e^{-\xi_1} =  a_2 + \xi_1 a_2  + {\cal O}(\xi^3) \;. 
\ee
Evaluating (\ref{CStep}) with the help of this relation 
we find
\be
a_{12} \ = \ a_1 + a_2 + {\xi_1} a_2 
-{1\over 2} a_1 a_2 -{1\over 2} {\xi_1} a_2  
+ {1\over 2} a_2 a_1   + {1\over 2} {\xi_2} a_1 \;, 
\ee
and we recover precisely (\ref{C}), completing 
the proof to second order.

\medskip
In the generalized setting, 
 the transformation of a gauge field is now implemented by
 ${\cal F}$, instead of ${\cal G}$.  Our expansion of ${\cal F}$
 to quadratic order is read 
 off from (\ref{Ftocubic}):  
 \be
 {\cal F} (\xi_1)  \ = \  1 + k_1   - {1\over 2} {\xi_1}
 k_1  + {1\over 2} k_1 k_1 \,, \quad ~~\hbox{with}
 ~~~~k_1 \equiv a_1
 -a_1^t\,, ~~~ a_1 \equiv \partial \xi_1\,.
 \ee 
This time the composition rule requires that
\be
{\cal F} (\xi_2) {\cal F}(\xi_1) \ = \ {\cal F} (\xi_{12}^c)\,. 
\ee
Written out to quadratic order it gives the requirement
 \be
 \label{dlk}
  {\bf 1} +k_1 + k_2 -\frac{1}{2}{\xi_1}k_1
  -\frac{1}{2}{\xi_1}k_2
  +k_2 k_1+\frac{1}{2}(k_1k_1+k_2k_2) \ = \ 
  {\bf 1} +k_{12}^c-\frac{1}{2}{\xi_{12}}k_{12}^c
  +\frac{1}{2}k_{12}^c k_{12}^c \;. 
 \ee
This equation will be satisfied if $\xi_{12}^c$ is such that
 \be\label{newKC}
  k_{12}^c \ = \ k_1+ k_2 -\frac{1}{2}[k_1, k_2]
  +\frac{1}{2}({\xi_1}k_2+{\xi_2}k_1 )\;.
 \ee
It remains to show that this is consistent with 
 \be
 \begin{split}
 ( \xi_{12}^c)^{M} \ = &\ \ \xi_1^{M}+\xi_2^{M}-\frac{1}{2}\big[\xi_1,\xi_2\big]_{\rm C}^{M}\\
  = &\ \ \xi_1^{M}+\xi_2^{M}-\frac{1}{2}\big[\xi_1,\xi_2\big]^{M}
  + \frac{1}{4}\xi_{1P}\partial^{M}\xi_2^{P}-\frac{1}{4}\xi_{2P}\partial^{M}\xi_1^{P}\;, 
\end{split}
 \ee
where we used (\ref{Cbracket}). 
The new $a_{12}^c$ here   is equal to the old $a_{12}$ in 
(\ref{C}),  
plus
the contributions from the last two terms above,      
 \be
  (a_{12}^c)_M{}^N \ = \ \partial_M\xi_{12}^N \ = \ 
  (a_{12})_M{}^N  + \frac{1}{4}\partial_M\big(\xi_{1P}\partial^{N}\xi_2^{P}-\xi_{2P}\partial^{N}\xi_1^{P}\big)\;.
 \ee
Therefore, 
 \be
 a_{12}^c \ = \ a_{12} + \frac{1}{4}\,(a_1a_2^t- a_2 a_1^t )
  + {1\over 4}  \bigl(\x_{1\,P}  (\partial \partial) \xi_2^P
  - \x_{2\,P}  (\partial \partial) \xi_1^P\bigr) \;.
 \ee
In the last couple of terms the matrix indices
are carried by the partial derivatives.  Now, when we form
$k_{12}^c = a_{12}^c - (a_{12}^c)^t$ those terms
cancel and we find
 \be\label{KCStep}
k_{12}^c \ = \ a_{12}^c- (a_{12}^c)^t \ = a_{12} -  a_{12}^t
+\frac{1}{2}(a_1 a_2^t- a_2 a_1^t)\;.
 \ee
Let us now expand the right-hand side of (\ref{newKC}) to see
if it agrees with the above $k_{12}^c$: 
 \be
 \label{grindit}
  \begin{split}
   k_1+\ &  k_2 -\frac{1}{2}[k_1, k_2]
  +\frac{1}{2}({\xi_1}k_2+{\xi_2}k_1 ) \\
  &\ = \ 
  a_1- a_1^t + a_2- a_2^t
  -\frac{1}{2}[a_1- a_1^t, a_2- a_2^t] 
  +\frac{1}{2}{\xi_1}(a_2- a_2^t)
  +\frac{1}{2}{\xi_2}(a_1-a_1^t) \\
  &\ = \ \   a_1  + a_2 - \frac{1}{2} [a_1, a_2]
  +\frac{1}{2}({\xi_1}a_2 +{\xi_2}a_1) \\
  &\qquad 
  - \Bigl( \,a_1^t   
  +a_2^t-\frac{1}{2}[a_2^t, a_1^t]+\frac{1}{2}({\xi_1}a_2^t+{\xi_2}a_1^t\big) \, \Bigr) 
  +\frac{1}{2}[a_1,a_2^t]+\frac{1}{2}[a_1^t,a_2] \\
  & \ = \ a_{12} - a_{12}^t
  +\frac{1}{2}(a_1a_2^t- a_2^ta_1 +a_1^ta_2 -
  a_2 a_1^t)\;, 
 \end{split}
\ee  
where we made use of (\ref{C}) to identify the terms that
comprise $a_{12}$ and $a_{12}^t$.   
We now note that 
\be
(a_1^t a_2)_{PQ} \ = \  (a_1)^M{}_P \,(a_2)_M{}^Q  \ = \ 
\partial^M \xi_{1P} \, \partial'_M \xi_2^Q  \ = \ 0 \,,
\ee
using the strong constraint in the form (\ref{stronglemma}).
For the same reason $a_2^t a_1 =0$.  As a result the last
right-hand side in (\ref{grindit}) indeed equals $k_{12}^c$,
as given in (\ref{KCStep}).  This proves the desired result.

To conclude, we explain how the composition law for generalized
coordinate transformations is consistent with the large transformation
of a scalar density, as postulated in (\ref{densitytrans}).   The consistency requires that
\be
\label{detcheck}
\det  \Bigl({\partial X \over \partial X''} \Bigr)\Bigl|_{X'' = e^{-\xi_{12}} X} \ = \ 
\det  \Bigl({\partial X \over \partial X''} \Bigr)\Bigl|_{X'' = e^{-\Theta (\xi_{12}^c) } X}  \;.
\ee
On the left-hand side we have the composition of determinants computed directly
by matrix multiplication as if the generalized coordinate transformations composed
directly; on the right hand side we have the determinant of the true composite 
generalized
transformation.   To verify this equality we recall the general identity
\be
\det (1 + A)  \ = \  \exp \Bigl[  \hbox{tr} \bigl( A -{1\over 2} A^2 + {1\over 3} A^3  + \cdots \bigr) \Bigr] \,,
\ee 
and from (\ref{cbhuseful-alt-vmvs}), when $X' = e^{-\xi} X$,  
  \be
\label{cbhuseful-alt-vmvs-vm}
 \frac{\partial X}{\partial X'} \ = \
1+ A \,, ~~~ \hbox{with} ~~~ A \ = \  a - {1\over 2} \xi a +{1\over 2} a^2 
+  {1\over 6} (\xi^2 a  - 2 (\xi a)a  - a\xi a + a^3)  + {\cal O}(\xi^4)\,. 
\ee
Equation (\ref{detcheck}) holds if the change 
\be
\label{xi}
\xi^M \to \xi^M + \cdots \partial^M \cdots 
\ee
leaves 
the computation of the determinant invariant.  This is because $\xi_{12}$ and $\xi_{12}^c$ differ
by such terms, and $\Theta (\xi)$ differs from $\xi$ by such terms.  As we can see above, the determinant is 
expressed in terms of traces of $A, A^2 , A^3, \ldots$.   We see immediately that
$\hbox{tr}\, a = \partial \cdot \xi$ is invariant under (\ref{xi}).  So is $\hbox{tr}\, \xi a = 
\xi^M \partial_M \partial \cdot \xi$, and 
\be
\hbox{tr} \, a^2 \ = \  \partial_M \xi^N \, \partial_N \xi^M \ = \ 
\partial_M (\xi^N + \cdots \partial^N \cdots )  \, \partial_N (\xi^M+  \cdots \partial^M \cdots ) 
\ee
The general term in any power of $A$ is made of a sequence of $a$ factors and $\xi$ operators,
and their trace will be invariant under (\ref{xi}): 
\be
\hbox{tr}\, [ a a \ldots (\xi a) \ldots  a] \ = \ \partial_M \xi^N \partial_N \xi^P \partial_P \ldots  \xi^R (\xi^Q\partial_Q)
\partial_R \xi^S \,\partial_S \ldots    \xi^W\partial_W \xi^M \,, 
\ee
since each $\xi$ index must be contracted with a derivative index (there are no $a^t$'s in here).
All in all this makes it manifest that (\ref{detcheck}) holds and that our formula for the 
large transformation of a density is consistent.

\section{Conclusions and open questions} 

We have presented a proposal for finite gauge transformations
in double field theory.  These transformations arise, in this viewpoint, from something we call generalized coordinate transformations.
While in ordinary geometry a vector field transforms with one power of the matrix of derivatives of the coordinate maps, in double field theory
a vector $A_M(X)$ transforms as
\be
\label{gkiu-conc}
A'(X') \ = \ {\cal F}(X', X) A (X)\,,
\ee
with 
\be\label{triv-conc}
{\cal F}(X', X) \ = \ {1 \over 2} \Big( {\partial X\over \partial X'}
{\partial X'\over \partial X}^t + {\partial X'\over \partial X}^t{\partial X\over \partial X'}\Big) \;.
\ee
Apart from passing a number of consistency
conditions a key property of the above expression is its relation to finite gauge transformations
defined more directly through the exponentiation of generalized
Lie derivatives $\widehat{\cal L}_\xi$:
\be
\label{vm-conc}
A'(X) \ = \ e^{\widehat{\cal L}_\xi}  A(X)  \ = \ e^{\,\xi + k}  A\,, ~~
~~k= a - a^t\,, ~~~ a = \partial \xi \,. 
\ee
To establish that this transformation is equivalent to the
transformation (\ref{gkiu-conc}) we had to show
that there is a generalized coordinate transformation 
$X' = f_\xi(X)$ in terms of $\xi$ for which the evaluation of ${\cal F}$
results in
\be
\label{sg-conc}
{\cal F} (X', X) \ = \ e^{-\xi} \, e^{\,\xi + k} \, ,   
\ee
 for this indeed implies the equivalence of
  (\ref{gkiu-conc}) and (\ref{vm-conc}). 
One may have thought that the coordinate transformation
$X' = e^{-\xi} X$ would do the job, but it turns out that
this only leads to (\ref{sg-conc}) holding to order $\xi^2$.  The generalized
coordinate transformation can be somewhat more exotic
while preserving familiar results due to some flexibility
afforded by use of the strong constraint.
We showed that in fact  
\be
X' \ = \ e^{-\Theta(\xi)} X \,,   ~~~\hbox{with} ~~~
\Theta^M \ = \ \xi^M  + {1\over 12}  \, (\xi \xi^L) \partial^M \xi_L 
+ {\cal O}(\xi^5) \,, \ee    
leads to (\ref{sg-conc}) up to and including ${\cal O}(\xi^4)$ terms.
Note that $\Theta^M$ equals $\xi^M$ to leading order and that the
cubic correction is a vector whose index is carried by a derivative.
This correction affects the coordinate transformation but 
also results in $\Theta^M\partial_M = \xi^M\partial_M$ on fields
(but not on~$X$). It remains an open  
problem to show that
 there exists a $\Theta(\xi) $  
that implies (\ref{sg-conc})  to all orders in $\xi$.
  It would also be of interest to understand the geometrical
  role of~$\Theta$. 

Generalized Lie derivatives define a Lie algebra.  Indeed, 
we have \cite{Hohm:2010pp}
\be
\bigl[\widehat{\cal L}_{\xi_1}\,, \widehat{\cal L}_{\xi_2}\bigr]\ = \ 
- \widehat{\cal L}_{[\xi_1, \xi_2]_c} \,,
\ee
with $[\cdot , \cdot ]_c$ the C-bracket, and the Jacobi identity holds:
\be
  \big[\big[\widehat{\cal L}_{\xi_1},\widehat{\cal L}_{\xi_2}\big],\widehat{\cal L}_{\xi_3}\big]
 + \big[\big[\widehat{\cal L}_{\xi_2},\widehat{\cal L}_{\xi_3}\big],\widehat{\cal L}_{\xi_1}\big]
 + \big[\big[\widehat{\cal L}_{\xi_3},\widehat{\cal L}_{\xi_1}\big],\widehat{\cal L}_{\xi_2}\big]
 \ = \ 0\;.
 \ee
This happens because the C-bracket 
Jacobiator of $(\xi_1, \xi_2, \xi_3)$
is a trivial parameter and generalized Lie derivatives of trivial
parameters are zero.  For both of the above properties one must
use the strong constraint. It is then a direct consequence of (\ref{vm-conc}) that the finite
transformations form a {\em group}. 
The Baker-Campbell-Hausdorff 
formula 
allows us to combine exponentials to get  
\be
\label{wkdflkd-conc}
e^{\widehat{\cal L}_{\xi_1(X)} } e^{\widehat{\cal L}_{\xi_2(X)} } \ = \ e^{\widehat{\cal L}_{\, {\xi}^c (\xi_2, \xi_1)} } \,, 
\ee
where
 \be
 \label{sgxi12vgc-conc}
 \xi^c (\xi_2, \xi_1)  \ = \ \xi_2 + \xi_1 + {1\over 2}  \big[\xi_2, \xi_1 \big]_c 
 + {1\over 12} \bigl(  \big[\xi_2,  \big[\xi_2, \xi_1 \big]_c \big]_c +  \big[\xi_1,  \big[\xi_1, \xi_2 \big]_c \big]_c \bigr)
 + \ldots \,.
  \ee
The group associativity property is guaranteed to hold acting on fields, namely
\be
\label{wkdflkd-conc-mod}
\bigl( e^{\widehat{\cal L}_{\xi_1(X)} } e^{\widehat{\cal L}_{\xi_2(X)} } \bigr)e^{\widehat{\cal L}_{\xi_3(X)} } \ = \ e^{\widehat{\cal L}_{\xi_1(X)} } \bigl( e^{\widehat{\cal L}_{\xi_2(X)} } e^{\widehat{\cal L}_{\xi_3(X)} }\bigr)  \,.
\ee
This results in 
 \be
   \exp\big(\,\widehat{\cal L}_{\,\xi^c (  \xi_3 , \xi^c (\xi_2, \xi_1) )}\,\big) \ = \ 
   \exp\big(\,\widehat{\cal L}_{\,\xi^c (  \xi^c(\xi_3, \xi_2) , \xi_1  )}\,\big) \;, 
 \ee 
and implies that the parameters of the left-hand side and right-hand side are equal up to 
a trivial parameter that does not generate a Lie derivative. A short computation shows that, in fact, 
\be\label{Jacobiator}
\xi^c \bigl( \, \xi_3\,, \, \xi^c (\xi_2, \xi_1) \,\bigr) \ = \ \xi^c \bigl(\,  \xi^c(\xi_3, \xi_2) \,,\, \xi_1 \,\bigr)-
\frac{1}{6}J(\xi_1,\xi_2,\xi_3)+\cdots \,,  
\ee
where $J(\xi_1,\xi_2,\xi_3)=[\xi_1,[\xi_2,\xi_3]_c]_c+{\rm cycl.}$ is the C-bracket Jacobiator that 
indeed is a trivial parameter, see eq.~(8.29) in \cite{Hull:2009zb}. 

In terms of generalized coordinate transformations we have
two maps $m_1: X \to X'$ and $m_2: X' \to X''$, 
\be
\begin{split}
X' \ = \ & \  e^{-\Theta(\xi_1)(X)} X\,,  \\
X'' \ = \ & \  e^{-\Theta(\xi_2)(X')} X'\,.
\end{split}
\ee
We are now to find the relevant map $m_{21}: X \to X''$.  The direct composition map is {\em not} the one we get. It would lead to a parameter built from $\xi_2$ and $\xi_1$ and the Lie bracket, not the C-bracket. What we get is the map $m_{21}=m_2\star m_1$ defined by 
\be
X'' \ =  \  e^{-\Theta \bigl(
 \xi^c (\,\xi_2\,, \,\xi_1\,)(X)\bigr)} X\,.
\ee
It may seem
paradoxical  
 that the direct composition $m_2\circ m_1$ of maps does not define the
map relevant in double field theory,  
but this is
unavoidable and consistent.  
Is it possible to write the 
exotic composition law we have here in terms of the maps rather 
than in terms of 
the generating $\xi$ parameters?  Should  
the coordinates be viewed in a different way that makes
the composition law look more natural?  

The exotic composition rule has important consequences for 
associativity. 
Consider a third map $m_3: X''\rightarrow X'''$, 
 \be
  X''' \ = \ e^{-\Theta(\xi_3)(X'')}X''\;.
 \ee 
Given the three maps $m_1$, $m_2$ and $m_3$, we can form a map 
$X\rightarrow X'''$ in two different ways, 
 \be
  m_3\star( m_2\star m_1 )\;, \qquad  (m_3\star m_2)\star m_1 \;. 
 \ee 
The first map leads to 
 \be\label{firstmap}
  X''' \ =  \ \exp\big(- \Theta(\xi^c (  \xi_3 , \xi^c (\xi_2, \xi_1) ))\big)X\;, 
 \ee
and the second map leads to 
  \be\label{secondmap}
  X''' \ =  \ \exp\big(- \Theta(\xi^c (  \xi^c(\xi_3, \xi_2) , \xi_1  ))\big)X\;.  
 \ee  
Due to (\ref{Jacobiator}) the two maps above are not equal. Indeed,  
a trivial parameter like the Jacobiator 
contributes to the transformation of $X$, see e.g.~(\ref{xtheta-vmvbvb}). 
Let us stress that this phenomenon would occur also without the modification 
from $\xi$ to $\Theta$ and that, moreover, this modification does not compensate 
for the difference between (\ref{firstmap}) and (\ref{secondmap}). 
Therefore, even though the generalized coordinate transformations build a 
group when acting on fields, 
the composition rule $\star$ for coordinate maps does not form a group. 
In this respect we note that recently there have been proposals 
that in string theory 
there is a plausible role for string coordinates that
are non-commutative or even non-associative
\cite{Blumenhagen:2010hj,Lust:2010iy,Blumenhagen:2011ph,Mylonas:2012pg}, 
and it would be interesting to investigate if 
the unconventional group structure 
encountered here
can be naturally interpreted in that context.
These are important open
questions, and any progress could help us learn about the
underlying geometry of string theory.

In double field theory the strong constraint guarantees that,
at least locally, we may always rotate into a frame where the
fields depend only on half of the (doubled) coordinates. It
is not yet known how to construct a  non-trivial patching 
of local regions of the doubled manifold 
leading to more general `non-geometric' configurations. 
The notion of a `T-fold', for instance, is based on the idea that
 field configurations 
on overlaps can be glued with the use of T-duality transformations \cite{Hull:2004in}. 
In order to address questions of this type 
in double field theory we need a clear picture of the finite gauge 
transformations, and in this paper we hope to have taken a step
in this direction.

\section*{Acknowledgments}  
We would like to thank Martin Rocek for discussions and Matt Headrick for a Mathematica  program that
helped do the power series computations in this paper.
B.~Zwiebach thanks the Harvard University Physics Department for hospitality during the period this research
was completed.

This work is supported by the U.S. Department of Energy (DoE) under the cooperative
research agreement DE-FG02-05ER41360, the
DFG Transregional Collaborative Research Centre TRR 33
and the DFG cluster of excellence ``Origin and Structure of the Universe".

\appendix

\section{Modifying the parameterization of the diffeomorphism}     
   
The purpose of this section is to verify that $\Theta$, as given
in (\ref{Thetatoquartic}), is actually correct to quartic order.
That is, no quartic term is needed and in fact
\be
\label{Thetatoquarticop}
\begin{split}
\Theta^M \ = \ & \ \xi^M - \delta_3^M + {\cal O}(\xi^5) 
=  \ \xi^M  + {1\over 12}  \, (\xi \xi^L) \partial^M \xi_L 
+ {\cal O}(\xi^5) \,, 
\end{split}
\ee    
will be sufficient to guarantee that 
\be
\label{FTheta=E-}
{\cal F}_\Theta = {\cal E}(a-a^t)  + {\cal O}(\xi^5)\,.
\ee
We begin by considering the discrepancy $\Delta{\cal F}$ between
${\cal F}$ and ${\cal E} (a-a^t)$ to quartic order in $\xi$. 
We write
\be\label{cubicrticdiff-vm-vm}
   {\cal F}
 \ =  \ {\cal E}(a-a^t)  - \Delta{\cal F} \,,
 \ee
 where  $\Delta {\cal F}$ is calculated
 by expansion of (\ref{vmvmvm}) and was calculated to leading
 cubic order before.  This time we find
\be
\label{vmvgvmvg}
\begin{split}
\Delta {\cal F}\ = \ & \ {1\over 12} \Bigl( (\xi a) a^t  
+ a^2 a^t - a\xi a^t
- a (a^t)^2 \Bigr)\\
&\hskip-10pt -\frac{1}{24}\Big[\, 
  (\xi a)aa^t -aa^t\xi a^t 
 +(\xi^2 a)a^t  -a(\xi^2a^t)\Bigr]
   \\
&\hskip-10pt    -{1\over 12} \Bigl[ 
 \, a^2\xi a^t  -a(\xi a^t)a^t
   + a^2 (a^t)^2 
 -{1\over 2} a^3a^t  - {1\over 2} a(a^t)^3  \Big]\,.
\end{split}
\ee 
The first line contains the contributions cubic in $\xi$, while
the other two lines contain the contributions quartic in $\xi$. 
Recall the expression for $X'$ and that for $X'_\Theta$ in 
(\ref{xtheta-vmvbvb}) 
 \be
 \label{xtheta-vmvb}
 \begin{split}
  X^{\prime M}_\xi \ \equiv \ &  \ 
  X^{M}-\xi^{M}+ {1\over 2} \,\xi \xi^M 
  -{1\over 3!}  \xi^2 \xi^M+{\cal O}(\xi^4)\,, \\ 
  X^{\prime M}_\Theta 
 \ \equiv  \ &  \    X^{M}-\Theta^{M}+ {1\over 2}  \xi \Theta^M 
 -{1\over 3!}  \xi^2 \Theta^M + {1\over 4!} \xi^3\Theta^M  + 
  {\cal O}(\xi^5)\;. 
\end{split}
 \ee
Using 
(\ref{xtheta-vmvb}) we can  write the relation between the two $X$'s as 
\be
X'^M_{\Theta}  \ = \  X'^M_{\xi}   + \delta_3^M
+ \hat\delta_4^M + \ldots \,,  ~~~\hbox{with}~~
 \hat\delta_4^M   \ = \    - {1\over 2} \xi \delta_3^M  \,.
\ee
Now define, for $i=3,4$, the derivatives  
\be
\label{defdeltahat}
({\Delta}_3)_Q{}^M  \ = \ \ \partial_Q  \delta_3^{~M}\,, ~~~~~
({\Delta}_4)_Q{}^M  \ = \ \ \partial_Q  \hat\delta_4^{~M}\;. 
\ee
With this notation, 
\be
\label{ghgh}
{\partial X'_\Theta\over \partial X} \ = \ 
{\partial X'_\xi\over \partial X} 
+  \Delta_3+  \Delta_4\,.  
\ee
A short calculation shows that 
\be
  {\Delta}_4   \ = \ - {1\over 2} (\xi + a) {\Delta}_3 \;. 
\ee
Now we need a formula to find the inverse of the above coordinate
derivatives.
Given the matrix $M$ expanded in powers of $\xi$ as
\be
M \ = \ 1 + A_1 + A_2 + A_3 + A_4 + {\cal O} (\xi^5)\;, 
\ee
with matrix inverse $M^{-1}$, we find that for the perturbed matrix
\be
M' \ = \ M  +    \Delta A_3  +  \Delta A_4 + {\cal O} (\xi^5)\,,
\ee
the inverse matrix is given by 
\be
M'^{-1} \ = \ M^{-1}   - \Delta A_3  - \Delta A_4  + 
(\Delta A_3) A_1 + A_1 (\Delta A_3)+  {\cal O} (\xi^5)\,.
\ee
Applied to (\ref{ghgh}) this gives
\be
\label{ghghs}
\begin{split}
{\partial X\over \partial X'_\Theta} \ = &\ \ 
{\partial X\over \partial X'_\xi}  - \Delta_3 - \Delta_4+  ( \Delta_3 (-a)  + (-a)  \Delta_3 ) \\
\ = &\ \ 
{\partial X\over \partial X'_\xi}  - \Bigl(
 \Delta_3 +  \Delta_4   + \Delta_3 \,a  + a\Delta_3 \Bigr)\;.  \\
\end{split}
\ee
We then find that 
\be
\label{fthetafxi}
{\cal F}_{\Theta} \ =  \  \ {\cal F}_{\xi}   + \Delta \,, 
\ee
where
\be
\begin{split}
 \Delta \ = & \  \  \Delta_3^t -  \Delta_3 
\ +  \Delta_4^t - \Delta_4 % \\ & 
-  ( \Delta_3  a + a  \Delta_3) + {1\over 2} ( \Delta_3 a^t + a  \Delta_3^t  +  \Delta_3^t a + a^t  \Delta_3)\,. 
\end{split}
\ee
In this light we have from (\ref{fthetafxi}) and (\ref{cubicrticdiff-vm-vm}) 
\be
{\cal F}_{\Theta} \ =  \    {\cal E}(a-a^t)  - \Delta{\cal F}  +  \Delta\;. 
\ee
So in order to get
${\cal F}_\Theta = {\cal E}(a-a^t)$  we need  a $\Theta (\xi)$ 
for which
\be
\label{mustshowvm}
\Delta \ = \   \Delta{\cal F} \,. 
\ee

Let us now confirm that our choice for $\Theta$, defined 
by (\ref{Thetatoquarticop}) with  
\be
\delta_3 \ = \ -  {1\over 12}  \, (\xi \xi^L) \partial^M \xi_L  \,,  
\ee
indeed produces the desired result. 
The definition (\ref{defdeltahat}) gives 
\be\label{ourcase}
\Delta_3 \ = \ -{1\over 12}  \bigl( (\xi a)a^t  + a^2 a^t \bigr)  
-{1\over 12}  (\xi \xi^L)  \partial \partial  \xi_L \,, 
\ee
where the matrix indices on the last term
are carried by the partial derivatives $\partial \partial$.  Moreover, 
\be\label{ourcase}
\begin{split} 
{\Delta}_4 \ = \ &    - {1\over 2} (\xi + a) \Delta_3 
  \\
  \ = \ & 
  \ \frac{1}{24}\Big[
  (\xi a)aa^t + (\xi^2a)a^t+  (\xi a)(\xi a^t)  
 +2a(\xi a)a^t
  +a^2\xi a^t+a^3a^t\Big]
  \\
  & +\frac{1}{24}(\xi +a)\bigl((\xi \xi^{P})\partial\partial \xi_{P}\bigr)\;. 
\end{split}
\ee
Using the above we can calculate all the ingredients of $\Delta$, 
\be
\begin{split}
\Delta_3^t - \Delta_3  \ = \  & \ 
{1\over 12} \Bigl( (\xi a) a^t  + a^2 a^t - a\xi  a^t
- a (a^t)^2 \Bigr)\;, \\
   -  (\Delta_3  a + a \Delta_3) &+\  {1\over 2} (\Delta_3 a^t + a \Delta_3^t  
   +  \Delta_3^t a + a^t  \Delta_3)\;, \\ 
    \ = \ & \ \frac{1}{12}\big(a(\xi a)a^t+a^3a^t-a^2(a^t)^2\big)\\
   &\hskip-10pt
    -\frac{1}{24}\big((\xi a)(a^t)^2+a^2\xi a^t \big)
   +\frac{1}{24}\big( \, a(\xi \xi^{P})\partial\partial\xi_{P}
   -(\xi \xi^{P}) (\partial\partial\xi_P)a^t\big)\;, \\
  {\Delta}_4^t- {\Delta}_4 \ = \ &\ -\frac{1}{24}\Big[\, 
  (\xi a)aa^t -aa^t\xi a^t 
 +(\xi^2a)a^t  -a(\xi^2a^t)\Bigr]
  \\
  &~ 
 -{1\over 24} \Bigl[ 
 \, 2a(\xi a)a^t 
   -2a(\xi a^t)a^t
   +a^2\xi a^t
   -(\xi a)(a^t)^2
 +a^3a^t  -a(a^t)^3  \Big]
 \\ &~
 -\frac{1}{24}\big( a(\xi \xi^P)\partial\partial\xi_{P}
  - (\xi \xi^{P}) (\partial\partial\xi_{P})a^t\big)\;, 
 \end{split}
 \ee
where  $\xi^2$ terms on the last line cancelled
because the `matrix' $\partial\partial$ is symmetric.
Adding up the above to find $\Delta$ we get
\be
\begin{split}
 \Delta \ = \ & \ {1\over 12} \Bigl( (\xi a) a^t  
+ a^2 a^t - a\xi  a^t
- a (a^t)^2 \Bigr)\\
%second line
&\hskip-10pt -\frac{1}{24}\Big[\, 
  (\xi a)aa^t -aa^t\xi a^t 
 +(\xi^2a)a^t  -a(\xi^2a^t)\Bigr]
  \\
%third line
&\hskip-10pt
 +  \frac{1}{12}\Bigl[a(\xi a)a^t+a^3a^t-a^2(a^t)^2
  -{1\over 2} (\xi a)(a^t)^2-{1\over 2} a^2\xi a^t \Bigr]
   \\
% fourth line
&\hskip-10pt    -{1\over 12} \Bigl[ 
 \, a(\xi a)a^t 
   -a(\xi a^t)a^t
   +{1\over 2} a^2\xi a^t
   -{1\over 2}(\xi a)(a^t)^2
 +{1\over 2} a^3a^t  - {1\over 2} a(a^t)^3  \Big]\;. 
\end{split}
\ee  
Combining the last two lines we get
\be
\begin{split}
\Delta \ = \ & \ {1\over 12} \Bigl( (\xi a) a^t  
+ a^2 a^t - a\xi a^t
- a (a^t)^2 \Bigr)\\
%second line
&\hskip-10pt -\frac{1}{24}\Big[\, 
  (\xi a)aa^t -aa^t\xi a^t 
 +(\xi^2a)a^t  -a(\xi^2a^t)\Bigr]
   \\
% fourth line
&\hskip-10pt    -{1\over 12} \Bigl[ 
 \, a^2\xi a^t  -a(\xi a^t)a^t
   + a^2 (a^t)^2 
 -{1\over 2} a^3a^t  - {1\over 2} a(a^t)^3  \Big]\;. 
\end{split}
\ee  
This coincides exactly with $\Delta {\cal F}$ in (\ref{vmvgvmvg}).
Thus equation (\ref{mustshowvm}) holds and we have completed 
the verification that ${\cal F}_\Theta = {\cal E}(a-a^t)$ up to terms
quintic in $\xi$.

\end{document}